\documentstyle[12pt]{article}
\makeatletter

\newdimen \paperheight  % needed for calculation of bottom margin.
\newdimen\phoffset
\newdimen\pvoffset
\newdimen\ptextwidth
\newdimen\ptextheight
\newdimen\ptopmargin
\newdimen\poddsidemargin
\newdimen\pevensidemargin

\newdimen\lhoffset
\newdimen\lvoffset
\newdimen\ltextwidth
\newdimen\ltextheight
\newdimen\ltopmargin
\newdimen\loddsidemargin
\newdimen\levensidemargin

\def\set@portland@values{%
\phoffset\hoffset
\pvoffset\voffset
\ptextwidth\textwidth
\ptextheight\textheight
\ptopmargin\topmargin
\poddsidemargin\oddsidemargin
\pevensidemargin\evensidemargin
\lhoffset \z@
\lvoffset \z@
\ltextwidth \ptextheight
\ltextheight \ptextwidth
\loddsidemargin \paperheight
\advance \loddsidemargin by -\ptextheight
\advance \loddsidemargin by -\headsep
\advance \loddsidemargin by -\headheight
\advance \loddsidemargin by -\ptopmargin
\advance \loddsidemargin by -\pvoffset
\advance \loddsidemargin by -2in
\levensidemargin \loddsidemargin
\ltopmargin \poddsidemargin
\advance \ltopmargin by -\headheight
\advance \ltopmargin by -\headsep
\advance \ltopmargin by \phoffset
}

\def\portrait{\clearpage \message{ \string\portrait }%
    \hoffset\phoffset
    \voffset\pvoffset
    \textwidth\ptextwidth
    \textheight\ptextheight
         \@colht\textheight  \@colroom\textheight \vsize\textheight
         \columnwidth\textwidth \@clubpenalty\clubpenalty
         \if@twocolumn \advance\columnwidth -\columnsep
         \divide\columnwidth\tw@ \hsize\columnwidth \@firstcolumntrue
         \fi
         \hsize\columnwidth \linewidth\hsize
    \topmargin\ptopmargin
    \oddsidemargin\poddsidemargin
    \evensidemargin\pevensidemargin
    }

\def\endportrait{\clearpage \message{ \string\endportrait }}

\def\landscape{\clearpage \message{ \string\landscape }%
    \hoffset\lhoffset
    \voffset\lvoffset
    \textwidth\ltextwidth
    \textheight\ltextheight
         \@colht\textheight  \@colroom\textheight \vsize\textheight
         \columnwidth\textwidth \@clubpenalty\clubpenalty
         \if@twocolumn \advance\columnwidth -\columnsep
         \divide\columnwidth\tw@ \hsize\columnwidth \@firstcolumntrue
         \fi
         \hsize\columnwidth \linewidth\hsize
    \topmargin\ltopmargin
    \oddsidemargin\loddsidemargin
    \evensidemargin\levensidemargin
    }

\def\endlandscape{\clearpage \message{ \string\endlandscape }}

\let\set@document@values \document
\def\document{\set@document@values \set@portland@values}

\newif\if@topcaption \@topcaptiontrue
\def\topcaption{\@topcaptiontrue\tablecaption}
\def\bottomcaption{\@topcaptionfalse\tablecaption}
\long\def\tablecaption{\refstepcounter{table} \@dblarg{\@xtablecaption}}
\long\def\@xtablecaption[#1]#2{%
  \long\def\@process@tablecaption{\@stcaption{table}[#1]{#2}}}
\let\@process@tablecaption\relax

\long\def\@stcaption#1[#2]#3{\par%
    \addcontentsline{\csname ext@#1\endcsname}{#1}%
        {\protect\numberline{\csname the#1\endcsname}{\ignorespaces #2}}
  \begingroup
    \@parboxrestore
    \normalsize
    \if@topcaption \vskip -10pt \fi % 'fix'
    \@makecaption{\csname fnum@#1\endcsname}{\ignorespaces #3}\par
    \if@topcaption \vskip 10pt \fi % 'fix'
  \endgroup}

\def\@tablehead{}
\def\tablehead#1{\gdef\@tablehead{#1}}
\def\tablefirsthead#1{\gdef\@table@first@head{#1}}

\def\@tabletail{}
\def\tabletail#1{%
    \gdef\@tabletail{\noalign{\global\let\@savcr=\\\global\let\\=\cr}%
                     #1\noalign{\global\let\\=\@savcr}}}
\def\tablelasttail#1{\gdef\@table@last@tail{#1}}

\newdimen\maxsize            % maximum pagesize
\newdimen\actsize            % actual pagesize
\newdimen\twocolsize         % needed for correct max height if twocolumn
\newdimen\parboxheight       % height plus depth of a parbox-argument
\newdimen\addspace           % stores the value of \\[#1]
\newdimen\midlineheight      % estimated size of a normal line
\newdimen\pargcorrection     % to set page height tolerance if p-arg
\newdimen\computedimens      % computation variable
\newbox\tabparbox

\def\@stabularcr{{\ifnum0=`}\fi\@ifstar{\@sxtabularcr}{\@sxtabularcr}}
\def\@sxtabularcr{\@ifnextchar[{\@sargtabularcr}%
                 {\ifnum0=`{\fi}\cr\nextline}}

\def\@sargtabularcr[#1]{\ifnum0=`{\fi}\ifdim #1>\z@
    \unskip\@sxargarraycr{#1}\else \@syargarraycr{#1}\fi}

\def\@sxargarraycr#1{\@tempdima #1\advance\@tempdima \dp \@arstrutbox%
    \vrule \@height\z@ \@depth\@tempdima \@width\z@ \cr%
    \noalign{\global\addspace=#1}\nextline}

\def\@syargarraycr#1{\cr\noalign{\vskip #1\global\addspace=#1}\nextline}

\def\@sstartpbox#1{\global\advance\maxsize by -\pargcorrection
                   \global\pargcorrection=0pt
                   \setbox\tabparbox%
                          \vtop\bgroup\hsize#1\@arrayparboxrestore}

\def\@sendpbox{\par\vskip\dp\@arstrutbox\egroup%
               \computedimens=\ht\tabparbox%
               \advance\computedimens by \dp\tabparbox%
               \ifnum\parboxheight<\computedimens
                  \global\parboxheight=\computedimens
               \fi
               \computedimens=0pt
               \box\tabparbox\hfil}

\def\calmidlineheight{\midlineheight=\arraystretch \baslineskp
                      \global\advance\midlineheight by 1pt
                      \global\pargcorrection=4\midlineheight}

\def\calpage{\global\actsize=\pagetotal  % where am I on the actual page?
             \twocolsize=\textheight            %  added 06.06.89
             \advance\twocolsize by -\@colroom  %        "
             \advance\actsize by \twocolsize    %        "
             \global\advance\actsize by \midlineheight
             \maxsize=\textheight        % start a new page when 90% of
             \multiply \maxsize by 9     % the page are used
             \divide\maxsize by 10
             \ifnum\actsize > \maxsize
                   \clearpage
                   \global\actsize=\pagetotal
             \fi
             \maxsize=\textheight       % now set \maxsize with tolerance
             \global\advance\maxsize by -\midlineheight}   % of one lines
\def\supertabular#1 {%           % before it was \edef\tableformat,
    \def\tableformat{\string#1} % store preamble
    \global\starfalse % remember this is the normal version

    \if@topcaption\@process@tablecaption
    \fi

    \def\baslineskp{\baselineskip}
    \calmidlineheight% estimate height of a normal line
    \calpage         % calculate max. pagesize and startpoint

    \let\@@tabularcr\@tabularcr%             Added JB 4/2/91
    \let\@tabularcr\@stabularcr
    \global\let\@oldcr=\\

    \let\@@startpbox=\@sstartpbox
    \let\@@endpbox=\@sendpbox
    \ifx\@table@first@head\undefined
        \let\@@tablehead=\@tablehead
    \else
        \let\@@tablehead=\@table@first@head
    \fi%                                     Added JB 4/2/91
    \begin{tabular}{\tableformat}%
    \@@tablehead}%   Added JB 15/2/91

\def\endsupertabular{%
    \ifx\@table@last@tail\undefined%
        \@tabletail%
    \else%
        \@table@last@tail%
    \fi%                                     Added JB 4/2/91
    \end{tabular}
    \let\@tabularcr\@@tabularcr             % Added JB 4/2/91
    \if@topcaption
    \else
        \@process@tablecaption
        \@topcaptiontrue
    \fi
    \global\let\\=\@oldcr
    \let\@table@first@head\undefined        % For the next ocurrence
    \let\@table@last@tail\undefined         % of this environment
    \let\@process@tablecaption\relax
}

\newif\ifstar
\newdimen\tabularwidth
\@namedef{supertabular*}#1#2 {% modified JB (15.2.91)
    \def\tableformat{\string#2} % store preamble
    \tabularwidth=#1 % The total width of the tabular
    \global\startrue % remember this is the *-version

    \if@topcaption\@process@tablecaption\fi

    \def\baslineskp{\baselineskip}
    \calmidlineheight% estimate height of a normal line
    \calpage         % calculate max. pagesize and startpoint

    \let\@@tabularcr\@tabularcr%              Added JB 4/2/91
    \let\@tabularcr\@stabularcr%              Added JB 4/2/91
    \global\let\@oldcr=\\

    \let\@@startpbox=\@sstartpbox
    \let\@@endpbox=\@sendpbox
    \ifx\@table@first@head\undefined
        \let\@@tablehead\@tablehead
    \else
        \let\@@tablehead\@table@first@head
    \fi%                                     Added JB 4/2/91
    \begin{tabular*}{\tabularwidth}{\tableformat}%
    \@@tablehead}%
\@namedef{endsupertabular*}{%
    \ifx\@table@last@tail\undefined%
        \@tabletail%
    \else%
        \@table@last@tail%
    \fi%                                     Added JB 4/2/91
    \end{tabular*}
    \let\@tabularcr\@@tabularcr
    \if@topcaption
    \else
        \@process@tablecaption
        \@topcaptiontrue
    \fi
    \global\let\\=\@oldcr
    \let\@table@first@head\undefined        % For the next ocurrence
    \let\@table@last@tail\undefined         % of this environment
    \let\@process@tablecaption\relax}

\def\nextline{%           %%% algorithm to calculate the pagebreaks %%%
    \noalign{\ifnum\parboxheight=0
                   \advance\actsize by \midlineheight
                   \global\advance\actsize by \addspace
             \else
                   \global\advance\actsize by \parboxheight
                   \divide\parboxheight by 11\relax
                   \global\advance\actsize by \parboxheight%
                   \global\parboxheight=0pt
             \fi
             \global\addspace=0pt}%
    \ifnum\actsize<\maxsize
    \noalign{\global\let\next\@empty}
    \else
         \@tabletail
         \ifstar%                     % Added 16-10-90
           \end{tabular*}%
         \else%
           \end{tabular}%
         \fi
         \if@twocolumn%                        % added 10.05.89
            \if@firstcolumn%                   %
               \newpage%                       %
            \else%                             %
               \clearpage%                     %
            \fi%                               %
            \twocolsize=\textheight%           % added 06.06.89
            \advance\twocolsize by -\@colroom% %       "
            \global\actsize=\twocolsize%       %       "
         \else                                 %
            \clearpage                         %
            \global\actsize=\midlineheight%
         \fi                                   %
         \let\next\@tablehead%                 % Added 15.2.91
         \ifstar%                              % Added 16-10-90
           \begin{tabular*}{\tabularwidth}{\tableformat}%
         \else%
           \begin{tabular}{\tableformat}%
         \fi%
    \fi\next}%                                % Added \next 15.2.91

\def\citen{\protect\p@citen}

\def\p@citen#1{%
\edef\@tempa{\@ignspaftercomma,#1, \@end, }% ignore spaces in parameter list
\edef\@tempa{\expandafter\@ignendcommas\@tempa\@end}%
\if@filesw \immediate \write \@auxout {\string \citation {\@tempa}}\fi
\@tempcntb\m@ne    % \@tempcntb tracks higest number
\let\@h@ld\relax   % nothing held from list yet
\let\@citea\@empty % no punctuation preceding first
\let\@celt\over    % not expandable, but identifiable
\def\@cite@list{}% % empty list to start
\@for \@citeb:=\@tempa \do{\@make@cite@list}% make a sorted list of numbers
\@tempcnta\m@ne    % no previous number
\let\@celt\@compress@cite \@cite@list % output number list with compression
\@h@ld}% output anything held over

\begingroup \catcode`\*=7 % funny catcode for comparisons
\gdef\@make@cite@list{%
 \expandafter\let \expandafter\@B@citeB \csname b@\@citeb \endcsname
 \ifx\@B@citeB\relax % undefined: output ? and warning
    \@citea {\bf ?}\let\@citea\citepunct
    \@warning {Citation `\@citeb' on page \thepage\space undefined}%
 \else %  defined
    \if *\ifnum\z@<0\@B@citeB *\else @\fi % a positive number, put in list
       \@tempcnta\@B@citeB \relax
       \ifnum \@tempcnta>\@tempcntb % new highest, add to end (efficiently)
          \edef\@cite@list{\@cite@list \@celt{\@B@citeB}}%
          \@tempcntb\@tempcnta
       \else % arbitrary number: insert appropriately
          \edef\@cite@list{\expandafter\@sort@celt \@cite@list \@gobble @}%
       \fi
    \else % citation is not a number, output immediately
       \@citea \@B@citeB \let\@citea\citepunct
 \fi\fi}
\endgroup % restore * catcode

\def\@compress@cite#1{%  % This is executed for each number
  \advance\@tempcnta\@ne % Now \@tempcnta is one more than the previous number
  \ifnum #1=\@tempcnta   % Number follows previous--hold on to it
     \ifx\@h@ld\relax    % first pair of successives
        \edef \@h@ld{\@citea #1}%
     \else               % compressible list of successives
        \edef \@h@ld{\hbox{--}\penalty\@highpenalty #1}%
     \fi
  \else   %  non-successor -- dump what's held and do this one
     \@h@ld \@citea #1\let\@h@ld\relax
  \fi \@tempcnta#1\let\@citea\citepunct
}

\def\@sort@celt#1#2{\ifx \@celt #1% parameters are \@celt{num}
     \ifnum #2<\@tempcnta % number goes later in list
        \@celt{#2}%
        \expandafter\expandafter\expandafter\@sort@celt % continue
     \else % number goes here
        \@celt{\number\@tempcnta}\@celt{#2}% stop comparing
  \fi\fi}

\def\citepunct{,\penalty\@highpenalty\hskip.13emplus.1emminus.1em}%
\def\cite{\protect\p@cite}

\def\p@cite{\@ifnextchar [{\@tempswatrue\@citex}{\@tempswafalse\@citex[]}}

\def\@citex[#1]#2{\@cite{\p@citen{#2}}{#1}}%

\def\@cite#1#2{\leavevmode\unskip
  \ifnum\lastpenalty=\z@ \penalty\@highpenalty \fi % highpenalty before
  \ [{\multiply\@highpenalty 3 #1% % triple-highpenalties within list
      \if@tempswa,\penalty\@highpenalty\ #2\fi % and before note.
    }]\spacefactor\@m}

\def\@ignspaftercomma#1, {\ifx\@end#1\@empty\else
   #1,\expandafter\@ignspaftercomma\fi}
\def\@ignendcommas,#1,\@end{#1}

\@addtoreset{equation}{section}

\ifx\@Hxfloat\@Hundef\else\expandafter\endinput\fi
\let\@Hxfloat\@xfloat
\def\@xfloat#1[{\@ifnextchar{H}{\@HHfloat{#1}[}{\@Hxfloat{#1}[}}
\def\@HHfloat#1[H]{%
\expandafter\let\csname end#1\endcsname\end@Hfloat
\vskip\intextsep\vbox\bgroup\def\@captype{#1}\parindent\z@
\ignorespaces}
\def\end@Hfloat{\egroup\vskip \intextsep}
\makeatother
\def\boldsymbol{}
\def\bfe#1{{\bf e}\left[ #1 \right]}
\def\opensquare{\thicklines\raisebox{-8pt}{\framebox(20,20){}}}
\def\sector#1#2{\ {\scriptstyle #1}\hspace{1mm}
\mathop{\opensquare}\limits_{\raisebox{-1mm}{$\scriptstyle#2$}}\hspace{1mm}}

\def\phc{\vphantom{$\overline{\chi(\hat{\cal M}_G)}$}}

\begin{document}
\addtolength{\baselineskip}{.3mm}
\thispagestyle{empty}
\vspace{-1.5cm}

\begin{flushright}
 KEK-TH-409\\
%{KEK preprint} ???\\
August  1994
\end{flushright}
\vspace{5mm}

\begin{center}
  {\large Duality of Orbifoldized Elliptic Genera}
  \footnote{{\noindent To appear in the proceedings of the workshop,
      {\em Quantum Field Theory, Integrable Models and Beyond}, Yukawa
      Institute for Theoretical Physics, Kyoto University, 14-18
      February 1994.  } }\\[13mm]
  {\sc Toshiya Kawai}\\[3mm]
  {\it
    National Laboratory for High Energy Physics (KEK),\\[2mm] Tsukuba,
    Ibaraki 305, Japan} \\[7mm]
  {\sc Sung-Kil Yang}\footnote
  { Supported in part by Grant-in-Aid for Scientific Research on
    Priority Area 231 ``Infinite Analysis'', Japan Ministry of
    Education.}\\[2mm]
  {\it Institute of Physics, University of
    Tsukuba, \\[2mm] Ibaraki 305, Japan}\\[18mm]
\end{center}
\vspace{.5cm}
\begin{center}
{\sc Abstract}
\end{center}
\vspace{.5cm}
\noindent We discuss duality and mirror symmetry phenomena of
Landau-Ginzburg orbifolds considering their elliptic genera.  Under
the duality (or mirror) transform performed by orbifoldizing the
Landau-Ginzburg model via some discrete group of the superpotential we
observe that the roles of the untwisted and twisted sectors are
exchanged. As explicit evidence detailed orbifold data are presented
for $N=2$ minimal models, Arnold's exceptional singularities, $K3$
surfaces constructed from Arnold's singularities and Fermat
hypersurfaces.

%\bigskip
%{\noindent  To appear in the proceedings of the workshop,
%{\em Quantum field theory, integrable models and beyond},
%Yukawa Institute for Theoretical Physics, Kyoto University,
%14-18 February 1994.  }

\bigskip
\begin{flushleft}
{\tt hep-th/9408121}
\end{flushleft}

\newpage

\section{Introduction}

Landau-Ginzburg field theory approach to two-dimensional critical
phenomena uncovers qualitative physical properties behind the exact
algebraic description in terms of conformal field theories
\cite{rZamolodchikov,rLC,rKMS}.  When $N=2$ supersymmetry is
considered even quantitative results can be deduced in the framework
of the Landau-Ginzburg models \cite{rMartinec,rVW,rLVW}. The
non-renormalization theorem for the superpotential of $N=2$ models is
believed to be responsible for this miracle.  $N=2$ Landau-Ginzburg
descriptions have also proved to be efficient in constructing
superstring vacua through orbifoldizing the Landau-Ginzburg models
\cite{rVafa,rIV}. This fact is somewhat mysterious since the
Landau-Ginzburg models do not a priori possess the target space
interpretation while more conventional sigma models have.  There have
been some arguments attempting to clarify the connection between the
Landau-Ginzburg models and the sigma models with Calabi-Yau target
spaces \cite{rGepneri,rGVW,rCecotti}.

Recently a novel scheme has been proposed to understand the
Landau-Ginzburg/Calabi-Yau correspondence \cite{rWitteniii}. One
considers a $U(1)$ gauged Landau-Ginzburg model with the
Fayet-Iliopoulos $D$-term and the theta term.  The model contains
several chiral superfields, one of which, say $P$, plays the role of
an order parameter. The coefficient $r$ of the Fayet-Iliopoulos term
combined with that of the theta term, $\theta$ turns out to be a
complex variable $t=\sqrt{-1}r+\frac{\theta}{2\pi}$ parametrizing the
complexified K\"ahler cone. By tuning $t$ one finds two extremum
regimes. One regime $(r \ll 0)$ represents the Landau-Ginzburg phase
where $p$, the bosonic component of $P$, acquires the vacuum
expectation value $\langle p \rangle \not= 0$. The $U(1)$ symmetry
then breaks down to some discrete group. This discrete group is
employed to orbifoldize the Landau-Ginzburg model.  In the other
regime $(r\gg 0)$ we have $\langle p \rangle = 0$ and the bosonic
components of the chiral superfields other than $P$ are constrained to
take values on a hypersurface in a weighted projective space. Hence
this is the regime of the sigma model. Furthermore it is argued that
one can make an analytic continuation from the Landau-Ginzburg to
Calabi-Yau regimes and vice versa on the complex $t$-plane. This
picture was also confirmed from another independent point of view
\cite{rAGM}\ and has been extended to the $(0,2)$ case
\cite{rWitteniii,rDK}.

One more evidence of the above scheme can be obtained by considering
the elliptic genus \cite{rSW,rWittenii,rAKMW}, {\it i.e.\/} the index
of the right-moving supercharge.  It is expected that the elliptic
genus is independent of the parameter $t$ because of its topological
nature and thus should coincide in both the Calabi-Yau and
Landau-Ginzburg orbifold phases.  Calculations of the elliptic genera
of the Landau-Ginzburg models were initiated in \cite{rWitteni}\ where
the A-type $N=2$ minimal model is taken to explain the essential idea.
This was soon followed by several groups who have extended the
original idea in order to incorporate various $N=2$ models
\cite{rWitteni, rDY, rKYY,rDAY,rHen,rAYS,rBH, rKM,rNW,rK}. In
ref.\cite{rKYY} it was confirmed that the elliptic genus of an
appropriate Landau-Ginzburg orbifold takes the same form as that of
the sigma model with a Calabi-Yau target manifold thus with a good
agreement with the above expectation.

The orbifoldized elliptic genus is also an interesting arena to
consider the mirror symmetry of Calabi-Yau manifolds \cite{rMirror}\
and similar phenomena. In fact the investigation of mirror symmetry
via the elliptic genus has already been taken up in \cite{rBH}.  A
well-known procedure to construct mirror pairs is to orbifoldize the
Landau-Ginzburg model via various symmetry groups of the
superpotential \cite{rGP}.  To compute the elliptic genera of the
resulting Landau-Ginzburg orbifolds necessitates a slightly more
involved formula than for the most frequently studied case. Thus after
introducing {\it generic\/} Landau-Ginzburg orbifolds in sect.2 we
briefly summarize the formulas for their elliptic genera in sect.3.

In sect.4, which is the main part of this contribution, we study
mirror phenomena and their cousins by employing these formulas.  By
picking up typical examples we present detailed data which have
accumulated during our series of analyses of the elliptic genus.
Although these data may be a sort of objects usually to be suppressed
in the literature it is quite impressive to experience duality or
mirror phenomena through explicit data. We thus think it worth
publishing these data in a comprehensible manner. We also believe that
our presentation is in accordance with the editorial spirit of this
proceedings.

\section{$N=2$ Landau-Ginzburg model and its orbifolds}

We consider the Landau-Ginzburg model whose Lagrangian density is
given by
\begin{equation}
  \int d^2\theta d^2\bar\theta\,\sum_{i=1}^NX_i\bar X_i+ \int
  d^2\theta\, W(X_i)+ \int d^2\bar\theta\, \bar W(\bar X_i)\,,
\end{equation}
where the superpotential $W$ is a weighted homogeneous polynomial of
$N$ chiral superfields $X_1, \cdots, X_N$ with weights
$\omega_1, \cdots, \omega_N$,
\begin{equation}
  \lambda W(X_1,\cdots,X_N)= W(\lambda^{\omega_1}
  X_1,\cdots,\lambda^{\omega_N} X_N)\,.
\end{equation}
We assume that $W$ has an isolated critical point at the origin and
$\omega_i$'s are strictly positive rational numbers such that
$\omega_1,\ldots,\omega_N\le\frac{1}{2}$.  The infrared fixed point
theory is believed to be described by an $N=2$ superconformal field
theory with
\begin{equation}
  \hat c = \sum_{i=1}^N(1-2\omega_i)\,.
\label{centralcharge}
\end{equation}

In general $W$ is invariant under some discrete group $G\subset
GL(N,{\bf C})$ acting on ${}^t(X_1,\ldots,X_N)$ and one can consider
the orbifold theory with respect to $G$.  The resulting theory is
called the Landau-Ginzburg orbifold and will be denoted symbolically
as $W/\!/G$ in the following. We shall restrict ourselves to the case
where $G$ is abelian and their elements take the form
$diag(\bfe{\alpha_1\omega_1},\ldots, \bfe{\alpha_N\omega_N})$ where
$\alpha_i\in {\bf Z}$ and $\bfe{*}=\exp(2\pi\sqrt{-1}*)$.  One
distinguished example of such discrete groups that always exists for
any $W$ is the one generated by
$diag(\bfe{\omega_1},\ldots,\bfe{\omega_N})$.  We shall call this
group as the {\it principal discrete group\/} and denote it by $G_0$. The
Landau-Ginzburg orbifold $W/\!/G_0$ is a fundamental and the most
frequently studied case.

In a favorable situation ({\it i.e.\/} $\hat c \in {\bf Z}$ and $G
\supseteq G_0$) the Landau-Ginzburg orbifold $W/\!/G$ can be
interpreted as an `analytic continuation' of some $N=2$ sigma model
with its target space  smoothed.

\section{Elliptic genus}

Topological properties of a supersymmetric theory in two space-time
dimensions can be succinctly summarized by the elliptic genus
\cite{rSW,rWittenii,rAKMW}.  This quantity was recently refined so as
to incorporate $N=2$ theories and up until now various examples have
been computed \cite{rWitteni, rDY, rKYY,rDAY,rHen,rAYS,rBH,
  rKM,rNW,rK}.

The definition of the $N=2$ elliptic genus is
\cite{rWitteni}\footnote{ Throughout this paper we shall consider
  $(2,2)$ theories only.}
\begin{equation}
  Z(\tau,z)=\mathop{\rm Tr}(-1)^F y^{{(J^{\rm L})}_0}q^{{\cal H}^{\rm
      L}}{\bar q}^{{\cal H}^{\rm R}},\quad y=\bfe{z},\
  q=\bfe{\tau}\quad (\mathop{\rm Im} \tau>0)\,,
  \label{defofegenus}
\end{equation}
where ${(J^{\rm L,R})}_0$ are the left, right $U(1)$ charge operators
and ${\cal H}^{\rm L,R}$ are the left, right Hamiltonians.  We have set
$(-1)^F=\exp[-\pi\sqrt{-1}\{(J^{\rm L})_0-(J^{\rm R})_0\}]$.  As
usual, due to the right supersymmetry $Z(\tau,z)$ is $\bar q$
independent.

The basic properties of the elliptic genus \cite{rKYY}\ are the
modular invariance up to a prefactor
\begin{equation}
  Z \left( {a\tau +b \over c\tau +d}, {z \over c\tau +d} \right)=
  \bfe{ \frac{\hat c}{2}\frac{ c z^2}{ c\tau +d}} Z(\tau, z)\,,\quad
  \pmatrix{ a&b\cr c&d\cr} \in SL(2,{\bf Z})\,,
  \label{modular}
\end{equation}
and the double quasi-periodicity
\begin{equation}
  Z(\tau, z+\lambda \tau +\mu)=(-1)^{\hat c (\lambda +\mu)}
  \bfe{-\frac{\hat c}{2} (\lambda^2 \tau +2\lambda z) } Z(\tau,
  z)\,,\quad \lambda, \mu \in h{\bf Z}\,,
  \label{period}
\end{equation}
where $h$ is the least positive integer such that the $U(1)$ charge of
any chiral ring element multiplied by $h$  is an integer.
These two properties together with the `$\chi_y$-genus'
\cite{rHirzebruch}\  determined by
\begin{equation}
  \chi_y=y^{\frac{\hat c}{2}}\lim_{\tau\rightarrow i\infty} Z(\tau,0)\,,
\end{equation}
characterize the elliptic genus.

The elliptic genus of the Landau-Ginzburg model can be computed as
\begin{equation}
  Z[W](\tau,z)=\prod_{i=1}^N {\vartheta_1(\tau,(1-\omega_i) z) \over
    \vartheta_1(\tau,\omega_i z)}\,,
  \label{LGgenus}
\end{equation}
where
\begin{equation}
  \begin{array}{rcl}
    &\displaystyle \vartheta_1(\tau,z)&= \sqrt{-1} \sum\limits_{n\in{\bf
        Z}}(-1)^n q^{\frac{1}{2}(n-\frac{1}{2})^2}
    y^{n-\frac{1}{2}}\\[2mm] &&\displaystyle = \sqrt{-1}q^{\frac{1}{8}}
    y^{-\frac{1}{2}} \prod_{n=1}^\infty(1-q^n)
    (1-q^{n-1}y)(1-q^ny^{-1})\,,
  \end{array}
\end{equation}
is one of the Jacobi theta functions. Using the theta function
formulae it is easy to check that $Z[W](\tau,z)$ obeys (\ref{modular})
and (\ref{period}) with $\hat c$ given by (\ref{centralcharge}) and
$h$ being the smallest positive integer such that $\omega_ih\in{\bf
  Z}$ for all $1\le i\le N$.

The elliptic genus of the Landau-Ginzburg orbifold $W/\!/G$ is given
by \cite{rKYY,rDAY,rBH}
\begin{equation}
  Z[W/\!/G](\tau,z)=\frac{1}{\vert G\vert}\sum_{{\boldsymbol
      \alpha},{\boldsymbol \beta}\in G} \epsilon({\boldsymbol
    \alpha},{\boldsymbol \beta})\sector{{\boldsymbol
      \beta}}{{\boldsymbol \alpha}}(\tau,z)\,,
\label{eqegenus}
\end{equation}
where
\begin{equation}
  \epsilon({\boldsymbol\alpha},{\boldsymbol \beta})=\prod_{i=1}^N
  (-1)^{\alpha_i+\beta_i+\alpha_i\beta_i}\,,
\label{eqsign}
\end{equation}
and
\begin{eqnarray}
  \sector{{\boldsymbol \beta}}{{\boldsymbol
      \alpha}}(\tau,z)&=&\prod_{i=1}^N
  \bfe{\frac{1-2\omega_i}{2}\alpha_i\beta_i}
  \bfe{\frac{1-2\omega_i}{2}(\alpha_i^2\tau+2\alpha_i
    z)}\nonumber\\[2mm] &&\hspace{3cm}\times
  \frac{\vartheta_1(\tau,(1-\omega_i)(z+\alpha_i\tau+\beta_i))}
  {\vartheta_1(\tau,\omega_i(z+\alpha_i\tau+\beta_i))}\,.
\end{eqnarray}
Here we have made  $\alpha=(\alpha_1,\ldots,\alpha_N)$ represent for the
element $diag(\bfe{\alpha_1\omega_1},\ldots,\bfe{\alpha_N\omega_N})$
of $G$.  Correspondingly the $\chi_y$-genus takes the form
\begin{equation}
  \chi_y[W/\!/G]=y^{\frac{\hat c}{2}}\frac{(-1)^N}{\vert
    G\vert}\sum_{{\boldsymbol \alpha},{\boldsymbol \beta}\in G}
  \prod_{\stackrel{\scriptstyle i}{\omega_i\alpha_i\not\in{\bf Z}}}
  y^{-(\!(\omega_i\alpha_i)\!)} \prod_{\stackrel{\scriptstyle
      i}{\omega_i\alpha_i\in{\bf Z}}} \frac{\sin
    \pi\{(\omega_i-1)z+\omega_i\beta_i\}}
  {\sin\pi(\omega_iz+\omega_i\beta_i)}\,,
\label{chiygenus}
\end{equation}
where $(\!( *)\!)=*-[*]-\frac{1}{2}$.

The orbifoldized elliptic genus $Z[W/\!/G]$ obeys the same modular
transformation property and double quasi-periodicity as those of
$Z[W]$.  In addition, if $\hat c$ is an integer, the conditions for
the double quasi-periodicity $\lambda,\mu\in h{\bf Z}$ are relaxed to
$\lambda,\mu\in {\bf Z}$ and the orbifold theory has a chance to have
correspondence with an $N=2$ sigma model.

The Witten index $Z[W/\!/G](\tau,0)=\chi_{y=1}[W/\!/G]$ reads
\begin{equation}
  \frac{(-1)^N}{\vert G\vert}\sum_{{\boldsymbol \alpha},{\boldsymbol
      \beta}\in G} \prod_{\stackrel{\scriptstyle
      i}{\omega_i\alpha_i\in{\bf Z} \ {\rm and}\
    \omega_i\beta_i\in{\bf Z}}} \left(1-\frac{1}{\omega_i}\right)\,,
\end{equation}
and this reproduces the result of Roan \cite{rRoan}\ which is in turn
the extension of Vafa's formula \cite{rVafa}
\footnote{
Notice that in the formula of $Z[W/\!/G_0](\tau,z)$ given in
ref.\cite{rKYY} we took
$\epsilon(\alpha,\beta)=(-1)^{D(\alpha+\beta+\alpha\beta)}$ where $D$
is an integer such that $Dh\equiv \hat c h\pmod{2}$.  Eq. (\ref{eqsign})
 corresponds to the choice  $D=N$ which is possible
since $\hat c h=Nh-2\sum_i\omega_ih\equiv Nh\pmod{2}$.  If $\hat c$ is
an integer we can instead take $D=\hat c$ which leads to the original
Vafa's formula \cite{rVafa}\ of the Euler characteristic.}.

\section{Self-duality, strange duality and mirror symmetry}

In this section we restrict ourselves to Landau-Ginzburg orbifolds
$W/\!/G$ with either $G=\{id\}$ or $G \supseteq G_0$. We now wish to
present a variety of computational results for the following
phenomena:

{\em There are some cases in which the Landau-Ginzburg orbifold
  $W/\!/G$ has a partner $W^*/\!/G^*$ such that
\begin{itemize}
\item $W^*/\!/G^*$ has the same central charge $\hat c$ as $W/\!/G$.
\item
  \begin{equation}
    Z[W/\!/G](\tau,z)=\pm Z[W^*/\!/G^*](\tau,z)\,,
\label{dualeg}
\end{equation}
or equivalently
\begin{equation}
  \chi_y[W/\!/G]=\pm \chi_y[W^*/\!/G^*]\,.
\label{dualchiy}
\end{equation}
\item By going from $W/\!/G$ to $W^*/\!/G^*$ the roles of the
  untwisted sectors and twisted sectors are interchanged.
\end{itemize}}
Here we have to explain what we mean by untwisted and twisted sectors.
If $G=\{ id\}$ we have only untwisted sectors and no twisted sectors.
If $G \supseteq G_0$, by untwisted sectors we mean the ones with
respect to the subgroup $G_0$. Thus the number of the untwisted
sectors is equal to $\vert G \vert /\vert G_0 \vert$.

Conceptual understanding of these observations is still lacking and it
is certainly true that a mere consideration of elliptic genus does not
suffice and perhaps we have to view things from a broader perspective
(see sect.5 for discussions).  Nevertheless we hope that a relative
ease of computations and their explicitness make these results worth
presenting.

\subsection{Self-duality of  minimal models}

Our first example is the well-known self-duality of $N=2$ minimal
model.  The $N=2$ minimal model is in one to one correspondence with
the Landau-Ginzburg model with its superpotential given by one of
$ADE$ potentials  and its central charge is given by $\hat
c=1-\frac{2}{h}$ where $h$ is the Coxeter number of $ADE$.  If we take
$W^*=W$ and choose $G=\{id\}$ and $G^*=G_0$ then the above situation
is realized as we now see.  As mentioned the Landau-Ginzburg model
$W=W/\!/\{id\}$ has no twisted sectors and its $\chi_y$-genus is, as
well-known, given by
\begin{equation}
  \chi_y[W]=\sum_{i=1}^lt^{m_i-1}\,,
\end{equation}
where $t=y^{\frac{1}{h}}$ and $m_1,\ldots,m_l$ are the exponents of
$ADE$.  The $\chi_y$-genus of the Landau-Ginzburg orbifold
$W/\!/G_0$  can be decomposed  as
\begin{equation}
  \chi_y[W/\!/G_0]=\sum_{\alpha=0}^{h-1}\chi_y^\alpha[W/\!/G_0]\,,
\end{equation}
and each contribution  is  given as follows.
For $A_l$ we have
\begin{equation}
  \chi_y^\alpha[W/\!/G_0]=\left\{
\begin{array}{ll}
        0\,,&\alpha=0\\[1mm]
       -t^{l-\alpha}\,,& \alpha=1,\ldots,l\,.
     \end{array}
\right.
\end{equation}
For $D_l$ and
if $l$ is even, we have
\begin{equation}
 \chi_y^\alpha[W/\!/G_0]=\left\{
\begin{array}{ll}
        0\,,&\alpha=\mbox{even}\\[1mm]
       -t^{2l-3-\alpha}\,,& \alpha=\mbox{odd}, \not = l-1\\[1mm]
        -2t^{l-2}\,,& \alpha=l-1
     \end{array}
\right. \,,
\end{equation}
while if $l$ is odd
\begin{equation}
 \chi_y^\alpha[W/\!/G_0]=\left\{
\begin{array}{ll}
        0\,,&\alpha=\mbox{even}, \not = l-1\\[1mm]
       -t^{2l-3-\alpha}\,,& \alpha=\mbox{odd}\\[1mm]
        -t^{l-2}\,,& \alpha=l-1
     \end{array}
\right. \,.
\end{equation}
For the remaining cases we have
\bigskip

\begin{tabular}{ll}\hline
  &\vphantom{\bigg\vert}$\{-\chi_y^0[W/\!/G_0], -\chi_y^1[W/\!/G_0], \ldots,
  -\chi_y^{h-1}[W/\!/G_0]\}$\\ \hline\\[-2mm]
  $E_6$&$\{ 0,{t^{10}},0,0,{t^7},{t^6},0,{t^4},{t^3},0,0,1\}$ \\
  $E_7$&$\{ 0,{t^{16}},0,0,0,{t^{12}},0,{t^{10}},0,{t^8},0,%
{t^6},0,{t^4},0,0,0,1\}$
\\ $E_8$&$\{ 0,{t^{28}},0,0,0,0,0,{t^{22}},0,0,0,{t^{18}},0,%
{t^{16}},0,0,0,{t^{12}},0,{t^{10}},0,0,0,{t^6},0,0,0,0,0,1\}$\\[-2mm]
\\  \hline
\end{tabular}
\bigskip

To summarize we found that
\begin{equation}
  \chi_y^\alpha[W/\!/G_0]=-\mbox{mult}(\alpha)\, t^{h-1-\alpha}\,,
\end{equation}
where $\mbox{mult}(\alpha)$ is the multiplicity of
$\alpha$ appearing in the set of exponents $\{m_1,\ldots,m_l\}$.
Hence it follows that
\begin{equation}
  \chi_y[W]=-\chi_y[W/\!/G_0]\,,
\end{equation}
and the twisted and the untwisted sectors are interchanged (with minus signs)
between $W$ and $W/\!/G_0$.

\subsection{Arnold's strange duality in terms of  Landau-Ginzburg orbifolds}

Let $W$ be the potential corresponding to one of Arnold's 14
exceptional singularities and let $W^*$ denote its dual in the sense
of strange duality. (See Table 1.)  $W$ and $W^*$ share the same
Coxeter number $h$ and hence $\hat c=1+2/h$.  Take $G=\{ id\}$ and
$G^*=G_0^*$ where $G_0^*$ is the principal discrete group of $W^*$.
Comparing Table 1 and Table 2 we find that
\begin{equation}
  \chi_y[W]=-\chi_y[W^*/\!/G_0^*]\,,
\end{equation}
and the twisted and the untwisted sectors are interchanged (with minus
signs) between $W$ and $W^*/\!/G_0^*$.

\subsection{Landau-Ginzburg orbifolds corresponding to  $K3$ constructed from
  exceptional singularities}

Let $\tilde W(z_1,z_2,z_3)$ be the potential corresponding to one of
Arnold's 14 exceptional singularities and let $\tilde
W^*(z_1,z_2,z_3)$ denote its dual in the sense of strange duality.
Set $W=W(z_1,z_2,z_3,z_4)=\tilde W(z_1,z_2,z_3)+z_4^h$
and similarly for $W^*$.
 Then it is known that one
can construct the $K3$ surface as the resolution of
\begin{equation}
  \{ (z_1,\ldots,z_4)\in {\bf WP}_{\{d_1,d_2,d_3,1\}}^3 \mid
  W(z_1,z_2,z_3,z_4)=0\}\,.
\end{equation}
The Landau-Ginzburg orbifold $W/\!/G_0$ describes the analytic
continuation of the $N=2$ sigma model whose target space is the $K3$
surface.
The $\chi_y$-genus of the $K3$ surface is
\begin{equation}
\chi_y(K3)=2+20y+2y^2\,,
\end{equation}
and we find $\chi_y[W/\!/G_0]=\chi_y(K3)=\chi_y[W^*/\!/G_0^*]$.
 Let us denote the
$\alpha^{\rm th}$ twisted sector contribution to the
$\chi_y[W/\!/G_0]$ by $\chi_y^\alpha[W/\!/G_0]$.  Table 3 shows that
the contributions from the untwisted sector and those from the twisted
sectors are interchanged between $W/\!/G_0$  and  $W^*/\!/G_0^*$. To put
differently,
\begin{equation}
\chi^0_y[W/\!/G_0]+\chi_y^0[W^*/\!/G_0^*]=\chi_y(K3)\,.
\end{equation}
Thus we have seen that the partner of $W/\!/G_0$ is given by
$W^*/\!/G_0^*$.

We remark that subjects related to what has been presented in the
previous and present subsections were earlier discussed by Martinec
\cite{rMartinec}.

\subsection{Mirror symmetry}
Our last example is mirror symmetry considered by Greene and Plesser
\cite{rGP}.  We consider the family of superpotentials given by
\begin{equation}
  W=z_1^d+\cdots+z_d^d\,,\quad d=3,4,5
\end{equation}
and  take $W=W^*$.
We choose $G$ to satisfy
\begin{equation}
  {\bf Z}_d\simeq G_0\subseteq G \subseteq ({\bf Z}_d)^{d-1}.
\end{equation}
Apparently the  number of such $G$'s is $2^{d-2}$ and they are given by:
\bigskip
\smallskip

\begin{center}
  \begin{tabular}{|l|l|l|}\hline
    \multicolumn{3}{|c|}{$d=3$} \\ \hline
     $G$ & \hfil generators \hfil & \phc$\chi(\hat{\cal M}_G)$\\ \hline
    $G_0\simeq {\bf Z}_3$& $(1,1,1)$&$0$\\
     $G_1\simeq ({\bf Z}_3)^2$ & $(1,1,1), (0,1,2)$ &$0$\\ \hline
  \end{tabular}
\end{center}
\smallskip

\begin{center}
  \begin{tabular}{|l|l|l|}\hline
    \multicolumn{3}{|c|}{$d=4$} \\ \hline
     $G$ & \hfil generators \hfil& \phc$\chi(\hat{\cal M}_G)$\\ \hline
     $G_0\simeq{\bf Z}_4$& $(1,1,1,1)$&$24$\\
     $G_1\simeq({\bf Z}_4)^2$ & $(1,1,1,1),(0,0,1,3)$ &$24$\\
     $G_2\simeq({\bf Z}_4)^2$ & $(1,1,1,1),(0,1,1,2)$ &$24$\\
     $G_3\simeq({\bf Z}_4)^3$ & $(1,1,1,1),(0,0,1,3),(0,1,1,2)$ &$24$\\
     \hline
  \end{tabular}
\end{center}
\smallskip

\begin{center}
  \begin{tabular}{|l|l|l|}\hline
    \multicolumn{3}{|c|}{$d=5$} \\ \hline
     $G$& \hfil generators \hfil & \phc$\chi(\hat{\cal M}_G)$\\ \hline
     $G_0\simeq {\bf Z}_5$& $(1,1,1,1,1)$&$-200$\\
     $G_1\simeq ({\bf Z}_5)^2$ & $(1,1,1,1,1), (0,0,0,1,4)$ &$-88$\\
     $G_2\simeq ({\bf Z}_5)^2$ & $(1,1,1,1,1), (0,1,2,3,4)$ &$-40$\\
     $G_3\simeq ({\bf Z}_5)^2$ & $(1,1,1,1,1), (0,1,1,4,4)$ &$8$\\
     $G_4\simeq ({\bf Z}_5)^3$ & $(1,1,1,1,1), (0,1,1,4,4), (0,1,2,3,4)$
&$-8$\\
     $G_5\simeq ({\bf Z}_5)^3$ & $(1,1,1,1,1), (0,1,3,1,0), (0,1,1,0,3)$
&$40$\\
     $G_6\simeq ({\bf Z}_5)^3$ & $(1,1,1,1,1), (0,1,4,0,0), (0,3,0,1,1)$
&$88$\\
     $G_7\simeq ({\bf Z}_5)^4$ & $(1,1,1,1,1), (0,1,2,3,4), (0,1,1,4,4),
      (0,0,0,1,4)$&$200$ \\ \hline
  \end{tabular}
\end{center}

\bigskip
\smallskip

{\noindent If $G=G_k$} then we take  $G^*=G_{2^{d-2}-1-k}$. Note that
$\vert G\vert \vert G^* \vert= d^d$.
The Landau-Ginzburg orbifold $W/\!/G$ corresponds to the sigma model on
$\hat{\cal M}_G$ which is a  resolution of the orbifold
\begin{equation}
  {\cal M}_G=\{ (z_1,\ldots,z_d)\in {\bf CP}^{d-1} : W(z_1,\ldots,z_d)=0\}/
   (G/G_0)\,.
\label{fermat}
\end{equation}
The Euler characteristic of $\hat{\cal M}_G$ is related to the
$\chi_y$-genus by
\begin{equation}
  \chi(\hat{\cal M}_G)=(-1)^d\chi_{y=1}[W/\!/G]\,.
\end{equation}

By examining the data presented below we can confirm that the asserted
situation indeed occurs. However before seeing this let us explain how
to look at tables below.  The elements of $G_i$ are ordered from left
to right then from top to bottom in their tabulations. Notice that the
elements of $G_i$ corresponding to the untwisted sectors take the form
$(0,*,\ldots,*)$. The $\chi_y^\alpha[W/\!/G_i]$ are arrayed in the
same order as for the elements of $G_i$ and should again be read from
left to right then from top to bottom in their tabulations.  Thus for
example the first and second rows of the table of $d=5$, $G_4$
correspond respectively to $1+5y+5y^2+y^3,0,0,0,0$ and
$0,0,2y+2y^2,0,2y+2y^2$ in the table of $\chi_y^\alpha[W/\!/G_4]$.

Now let us consider, as an illustration, the pair of $W/\!/G_1$ and
$W/\!/G_2$ for $d=4$. Both theories have $4$ untwisted sectors and 12
twisted sectors.  The total twisted contribution to
$\chi_y[W/\!/G_1]$ reads $y^2 + 0 + y + 0 +y +y+ y+ y +1 + 0 + y +
0=1+6y+y^2$ while the total untwisted contributions to
$\chi_y[W/\!/G_2]$ reads $(1+5y+y^2)+0+y+0=1+6y+y^2$.  As another
example let us take the pair of $W/\!/G_0$ and $W/\!/G_7$ for $d=5$.
The total twisted contribution to $\chi_y[W/\!/G_0]$ reads
$-y^3-y^2-y-1$. The theory $W/\!/G_7$ has $5^4/5=125$ untwisted
sectors. The first one makes a contribution of $1+y+y^2+y^3$ while
each of the remaining $124$ ones of $0$.

The other cases can be checked similarly. Though we have not worked
out, it is also likely that similar results can be obtained for a
class of mirror pairs considered in \cite{rBeHu}.

%\vspace{2cm}
\newpage

\footnotesize
\def\phci{\vphantom{\Big\vert}}
\begin{center}
  (i) $d=3$
\end{center}
\bigskip

%\topcaption{}
\tablefirsthead{\hline \multicolumn{4}{|c|}{$G_0$ }\\ \hline}
%\tabletail{\hline \multicolumn{4}{r}{{\it cont. to next page }}\\ }
\tablelasttail{\hline}
\begin{center}
\begin{supertabular}{|llll|}
 &(0, 0, 0) & (1, 1, 1) & (2, 2, 2)\\
\end{supertabular}
\end{center}
\vspace{1cm}

%\topcaption{}
\tablefirsthead{\hline \multicolumn{4}{|c|}{$G_1$ }\\ \hline}
%\tabletail{\hline \multicolumn{4}{r}{{\it cont. to next page }}\\ }
\tablelasttail{\hline}
\begin{center}
\begin{supertabular}{|llll|}
 &(0, 0, 0) & (0, 1, 2) & (0, 2, 4)\\
 &(1, 1, 1) & (1, 2, 3) & (1, 3, 5)\\
 &(2, 2, 2) & (2, 3, 4) & (2, 4, 6)\\
\end{supertabular}
\end{center}
\vspace{1cm}

%\topcaption{Twisted sector contributions to $\chi_y[W/\!/G_0]$}
\tablefirsthead{\hline \multicolumn{1}{|c|}{$\chi_y^\alpha[W/\!/G_0]$
}\\ \hline}
%\tablehead{\hline}
%\tabletail{\hline\\}
\tablelasttail{\hline}
\begin{center}
\begin{supertabular}{|l|}
$ 1 + y,-y,-1$\\
\end{supertabular}
\end{center}
\vspace{2cm}

%\topcaption{Twisted sector contributions to $\chi_y$ of $\hat {\cal M}_{G_1}$}
\tablefirsthead{\hline \multicolumn{4}{|c|}{$\chi_y^\alpha[W/\!/G_1]$
}\\ \hline}
%\tablehead{\hline}
%\tabletail{\hline\\}
\tablelasttail{\hline}
\begin{center}
\begin{supertabular}{|llll|}
 &$1 + y,0,0,$
  &$ -y,0,0,$
 &$  -1,0,0 $\\
\end{supertabular}
\end{center}
\vspace{2cm}

\newpage

\begin{center}
 (ii)  $d=4$
\end{center}
\bigskip

%\topcaption{}
\tablefirsthead{\hline \multicolumn{5}{|c|}{$G_0$ }\\ \hline}
%\tabletail{\hline \multicolumn{5}{r}{{\it cont. to next page }}\\ }
\tablelasttail{\hline}
%%\topcaption{}
\begin{center}
\begin{supertabular}{|lllll|}
 &(0, 0, 0, 0) & (1, 1, 1, 1) & (2, 2, 2, 2) & (3, 3, 3, 3)\\
\end{supertabular}
\end{center}
\vspace{.5cm}

%\topcaption{}
\tablefirsthead{\hline \multicolumn{5}{|c|}{$G_1$ }\\ \hline}
%\tabletail{\hline \multicolumn{5}{r}{{\it cont. to next page }}\\ }
\tablelasttail{\hline}
%%\topcaption{}
\begin{center}
\begin{supertabular}{|lllll|}
&(0, 0, 0, 0)&(0, 0, 1, 3)&(0, 0, 2, 6)&(0, 0, 3, 9)\\
&(1, 1, 1, 1)&(1, 1, 2, 4)&(1, 1, 3, 7)&(1, 1, 4, 10)\\
&(2, 2, 2, 2)&(2, 2, 3, 5)&(2, 2, 4, 8)&(2, 2, 5, 11)\\
&(3, 3, 3, 3)&(3, 3, 4, 6)&(3, 3, 5, 9)&(3, 3, 6, 12)\\
\end{supertabular}
\end{center}
\vspace{.5cm}

%\topcaption{}
 \tablefirsthead{\hline \multicolumn{5}{|c|}{$G_2$ }\\ \hline}
%\tabletail{\hline \multicolumn{5}{r}{{\it cont. to next page }}\\ }
\tablelasttail{\hline}
%%\topcaption{}
\begin{center}
\begin{supertabular}{|lllll|}
&(0, 0, 0, 0)&(0, 1, 1, 2)&(0, 2, 2, 4)&(0, 3, 3, 6)\\
&(1, 1, 1, 1)&(1, 2, 2, 3)&(1, 3, 3, 5)&(1, 4, 4, 7)\\
&(2, 2, 2, 2)&(2, 3, 3, 4)&(2, 4, 4, 6)&(2, 5, 5, 8)\\
&(3, 3, 3, 3)&(3, 4, 4, 5)&(3, 5, 5, 7)&(3, 6, 6, 9)\\
\end{supertabular}
\end{center}
\vspace{.5cm}

%\topcaption{}
 \tablefirsthead{\hline \multicolumn{5}{|c|}{$G_3$ }\\ \hline}
%\tabletail{\hline \multicolumn{5}{r}{{\it cont. to next page }}\\ }
\tablelasttail{\hline}
%%\topcaption{}
\begin{center}
\begin{supertabular}{|lllll|}
&(0, 0, 0, 0)&(0, 1, 1, 2)&(0, 2, 2, 4)&(0, 3, 3, 6)\\
&(0, 0, 1, 3)&(0, 1, 2, 5)&(0, 2, 3, 7)&(0, 3, 4, 9)\\
&(0, 0, 2, 6)&(0, 1, 3, 8)&(0, 2, 4, 10)&(0, 3, 5, 12)\\
&(0, 0, 3, 9)&(0, 1, 4, 11)&(0, 2, 5, 13)&(0, 3, 6, 15)\\
&(1, 1, 1, 1)&(1, 2, 2, 3)&(1, 3, 3, 5)&(1, 4, 4, 7)\\
&(1, 1, 2, 4)&(1, 2, 3, 6)&(1, 3, 4, 8)&(1, 4, 5, 10)\\
&(1, 1, 3, 7)&(1, 2, 4, 9)&(1, 3, 5, 11)&(1, 4, 6, 13)\\
&(1, 1, 4, 10)&(1, 2, 5, 12)&(1, 3, 6, 14)&(1, 4, 7, 16)\\
&(2, 2, 2, 2)&(2, 3, 3, 4)&(2, 4, 4, 6)&(2, 5, 5, 8)\\
&(2, 2, 3, 5)&(2, 3, 4, 7)&(2, 4, 5, 9)&(2, 5, 6, 11)\\
&(2, 2, 4, 8)&(2, 3, 5, 10)&(2, 4, 6, 12)&(2, 5, 7, 14)\\
&(2, 2, 5, 11)&(2, 3, 6, 13)&(2, 4, 7, 15)&(2, 5, 8, 17)\\
&(3, 3, 3, 3)&(3, 4, 4, 5)&(3, 5, 5, 7)&(3, 6, 6, 9)\\
&(3, 3, 4, 6)&(3, 4, 5, 8)&(3, 5, 6, 10)&(3, 6, 7, 12)\\
&(3, 3, 5, 9)&(3, 4, 6, 11)&(3, 5, 7, 13)&(3, 6, 8, 15)\\
&(3, 3, 6, 12)&(3, 4, 7, 14)&(3, 5, 8, 16)&(3, 6, 9, 18)\\
\end{supertabular}
\end{center}
\newpage

%\topcaption{Twisted sector contributions to $\chi_y$ of ${\cal M}_{G_0}$}
\tablefirsthead{\hline \multicolumn{1}{|c|}{$\chi_y^\alpha[W/\!/G_0]$ }\\
\hline}
%\tablehead{\hline}
%\tabletail{\hline\\}
\tablelasttail{\hline}
\begin{center}
\begin{supertabular}{|l|}
$ 1 + 19y + {y^2},{y^2},y,1$\\
\end{supertabular}
\end{center}
\vspace{2cm}

%\topcaption{Twisted sector contributions to $\chi_y$ of ${\cal M}_{G_0}$}
\tablefirsthead{\hline \multicolumn{5}{|c|}{$\chi_y^\alpha[W/\!/G_1]$ }\\
\hline}
%\tablehead{\hline}
%\tabletail{\hline\\}
\tablelasttail{\hline}
\begin{center}
\begin{supertabular}{|lllll|}
&$1 + 5y + {y^2},3y,3y,3y,$
&$  {y^2},0,y,0,$
&$  y,y,y,y,$
&$   1,0,y,0$\\
\end{supertabular}
\end{center}
\vspace{2cm}

 %\topcaption{Twisted sector contributions to $\chi_y$ of ${\cal M}_{G_2}$}
\tablefirsthead{\hline \multicolumn{5}{|c|}{$\chi_y^\alpha[W/\!/G_2]$}\\
\hline}
%\tablehead{\hline}
%\tabletail{\hline\\}
\tablelasttail{\hline}
\begin{center}
\begin{supertabular}{|lllll|}
&$ 1 + 5y + {y^2},0,y,0,$
&$   {y^2},y,y,3y,$
&$   y,0,3y,0,$
&$   1,3y,y,y$\\
\end{supertabular}
\end{center}
\vspace{2cm}

%\topcaption{Twisted sector contributions to $\chi_y$ of ${\cal M}_{G_0}$}
\tablefirsthead{\hline \multicolumn{5}{|c|}{$\chi_y^\alpha[W/\!/G_3]$}\\
\hline}
%\tablehead{\hline}
%\tabletail{\hline\\}
\tablelasttail{\hline}
\begin{center}
\begin{supertabular}{|lllll|}
 & $ 1 + y + {y^2},0,0,0,$
 & $ 0,0,0,0,$
 & $ 0,0,0,0,$
 & $ 0,0,0,0,$\\
 & $ {y^2},y,y,0,$
 & $ 0,y,0,0,$
 & $ y,0,y,0,$
 & $ 0,0,y,0,$\\
 & $ y,0,0,0,$
 & $ y,0,0,y,$
 & $ 0,y,0,y,$
 & $ y,y,0,0,$\\
 & $ 1,0,y,y,$
 & $ 0,0,y,0,$
 & $ y,0,y,0,$
 & $ 0,0,0,y $\\
\end{supertabular}
\end{center}
\newpage

\begin{center}
  (iii) $d=5$
\end{center}
\bigskip

%% quintic %%
%\topcaption{}
\tablefirsthead{\hline \multicolumn{6}{|c|}{$G_0$ }\\ \hline}
%\tabletail{\hline \multicolumn{6}{r}{{\it cont. to next page }}\\ }
\tablelasttail{\hline}
%%\topcaption{The elements of $G_0$}
\begin{center}
\begin{supertabular}{|llllll|}
  &(0, 0, 0, 0, 0) & (1, 1, 1, 1, 1) & (2, 2, 2, 2, 2) & (3, 3, 3, 3,
  3) & (4, 4, 4, 4, 4)\\
\end{supertabular}
\end{center}
\vspace{2cm}

%\topcaption{}
\tablefirsthead{\hline \multicolumn{6}{|c|}{$G_1$ }\\ \hline}
%\tabletail{\hline \multicolumn{6}{r}{{\it cont. to next page }}\\ }
\tablelasttail{\hline}
%%\topcaption{The elements of $G_1$}
\begin{center}
\begin{supertabular}{|llllll|}
  &(0, 0, 0, 0, 0) & (0, 0, 0, 1, 4) & (0, 0, 0, 2, 8) & (0, 0, 0, 3,
  12) & (0, 0, 0, 4, 16)\\ &(1, 1, 1, 1, 1) & (1, 1, 1, 2, 5) & (1, 1,
  1, 3, 9) & (1, 1, 1, 4, 13) & (1, 1, 1, 5, 17)\\ &(2, 2, 2, 2, 2) &
  (2, 2, 2, 3, 6) & (2, 2, 2, 4, 10) & (2, 2, 2, 5, 14) & (2, 2, 2, 6,
  18)\\ &(3, 3, 3, 3, 3) & (3, 3, 3, 4, 7) & (3, 3, 3, 5, 11) & (3, 3,
  3, 6, 15) & (3, 3, 3, 7, 19)\\ &(4, 4, 4, 4, 4) & (4, 4, 4, 5, 8) &
  (4, 4, 4, 6, 12) & (4, 4, 4, 7, 16) & (4, 4, 4, 8, 20)\\
\end{supertabular}
\end{center}
\vspace{2cm}

%\topcaption{}
\tablefirsthead{\hline \multicolumn{6}{|c|}{$G_2$ }\\ \hline}
%\tabletail{\hline \multicolumn{6}{r}{{\it cont. to next page }}\\ }
\tablelasttail{\hline}
%%\topcaption{The elements of $G_2$}
\begin{center}
\begin{supertabular}{|llllll|}
  &(0, 0, 0, 0, 0) & (0, 1, 2, 3, 4) & (0, 2, 4, 6, 8) & (0, 3, 6, 9,
  12) & (0, 4, 8, 12, 16)\\ &(1, 1, 1, 1, 1) & (1, 2, 3, 4, 5) & (1,
  3, 5, 7, 9) & (1, 4, 7, 10, 13) & (1, 5, 9, 13, 17)\\ &(2, 2, 2, 2,
  2) & (2, 3, 4, 5, 6) & (2, 4, 6, 8, 10) & (2, 5, 8, 11, 14) & (2, 6,
  10, 14, 18)\\ &(3, 3, 3, 3, 3) & (3, 4, 5, 6, 7) & (3, 5, 7, 9, 11)
  & (3, 6, 9, 12, 15) & (3, 7, 11, 15, 19)\\ &(4, 4, 4, 4, 4) & (4, 5,
  6, 7, 8) & (4, 6, 8, 10, 12) & (4, 7, 10, 13, 16) & (4, 8, 12, 16,
  20)\\
\end{supertabular}
\end{center}
\vspace{2cm}

%\topcaption{}
\tablefirsthead{\hline \multicolumn{6}{|c|}{$G_3$ }\\ \hline}
%\tabletail{\hline \multicolumn{6}{r}{{\it cont. to next page }}\\ }
\tablelasttail{\hline}
%%\topcaption{The elements of $G_3$}
\begin{center}
\begin{supertabular}{|llllll|}
  &(0, 0, 0, 0, 0) & (0, 1, 1, 4, 4) & (0, 2, 2, 8, 8) & (0, 3, 3, 12,
  12) & (0, 4, 4, 16, 16)\\ &(1, 1, 1, 1, 1) & (1, 2, 2, 5, 5) & (1,
  3, 3, 9, 9) & (1, 4, 4, 13, 13) & (1, 5, 5, 17, 17)\\ &(2, 2, 2, 2,
  2) & (2, 3, 3, 6, 6) & (2, 4, 4, 10, 10) & (2, 5, 5, 14, 14) & (2,
  6, 6, 18, 18)\\ &(3, 3, 3, 3, 3) & (3, 4, 4, 7, 7) & (3, 5, 5, 11,
  11) & (3, 6, 6, 15, 15) & (3, 7, 7, 19, 19)\\ &(4, 4, 4, 4, 4) & (4,
  5, 5, 8, 8) & (4, 6, 6, 12, 12) & (4, 7, 7, 16, 16) & (4, 8, 8, 20,
  20)\\
\end{supertabular}
\end{center}

\newpage

%\topcaption{}
\tablefirsthead{\hline \multicolumn{6}{|c|}{$G_4$ }\\ \hline}
%\tablehead{\multicolumn{6}{l}{{\it cont. from previous page }}\\
%  \hline}
%\tabletail{\hline \multicolumn{6}{r}{{\it cont. to next page }}\\ }
\tablelasttail{\hline}
%%\topcaption{The elements of $G_4$}
\begin{center}
\begin{supertabular}{|llllll|}
  &(0, 0, 0, 0, 0) & (0, 1, 2, 3, 4) & (0, 2, 4, 6, 8) & (0, 3, 6, 9,
  12) & (0, 4, 8, 12, 16)\\ &(0, 1, 1, 4, 4) & (0, 2, 3, 7, 8) & (0,
  3, 5, 10, 12) & (0, 4, 7, 13, 16) & (0, 5, 9, 16, 20)\\ &(0, 2, 2,
  8, 8) & (0, 3, 4, 11, 12) & (0, 4, 6, 14, 16) & (0, 5, 8, 17, 20) &
  (0, 6, 10, 20, 24)\\ &(0, 3, 3, 12, 12) & (0, 4, 5, 15, 16) & (0, 5,
  7, 18, 20) & (0, 6, 9, 21, 24) & (0, 7, 11, 24, 28)\\ &(0, 4, 4, 16,
  16) & (0, 5, 6, 19, 20) & (0, 6, 8, 22, 24) & (0, 7, 10, 25, 28) &
  (0, 8, 12, 28, 32)\\ &(1, 1, 1, 1, 1) & (1, 2, 3, 4, 5) & (1, 3, 5,
  7, 9) & (1, 4, 7, 10, 13) & (1, 5, 9, 13, 17)\\ &(1, 2, 2, 5, 5) &
  (1, 3, 4, 8, 9) & (1, 4, 6, 11, 13) & (1, 5, 8, 14, 17) & (1, 6, 10,
  17, 21)\\ &(1, 3, 3, 9, 9) & (1, 4, 5, 12, 13) & (1, 5, 7, 15, 17) &
  (1, 6, 9, 18, 21) & (1, 7, 11, 21, 25)\\ &(1, 4, 4, 13, 13) & (1, 5,
  6, 16, 17) & (1, 6, 8, 19, 21) & (1, 7, 10, 22, 25) & (1, 8, 12, 25,
  29)\\ &(1, 5, 5, 17, 17) & (1, 6, 7, 20, 21) & (1, 7, 9, 23, 25) &
  (1, 8, 11, 26, 29) & (1, 9, 13, 29, 33)\\ &(2, 2, 2, 2, 2) & (2, 3,
  4, 5, 6) & (2, 4, 6, 8, 10) & (2, 5, 8, 11, 14) & (2, 6, 10, 14,
  18)\\ &(2, 3, 3, 6, 6) & (2, 4, 5, 9, 10) & (2, 5, 7, 12, 14) & (2,
  6, 9, 15, 18) & (2, 7, 11, 18, 22)\\ &(2, 4, 4, 10, 10) & (2, 5, 6,
  13, 14) & (2, 6, 8, 16, 18) & (2, 7, 10, 19, 22) & (2, 8, 12, 22,
  26)\\ &(2, 5, 5, 14, 14) & (2, 6, 7, 17, 18) & (2, 7, 9, 20, 22) &
  (2, 8, 11, 23, 26) & (2, 9, 13, 26, 30)\\ &(2, 6, 6, 18, 18) & (2,
  7, 8, 21, 22) & (2, 8, 10, 24, 26) & (2, 9, 12, 27, 30) & (2, 10,
  14, 30, 34)\\ &(3, 3, 3, 3, 3) & (3, 4, 5, 6, 7) & (3, 5, 7, 9, 11)
  & (3, 6, 9, 12, 15) & (3, 7, 11, 15, 19)\\ &(3, 4, 4, 7, 7) & (3, 5,
  6, 10, 11) & (3, 6, 8, 13, 15) & (3, 7, 10, 16, 19) & (3, 8, 12, 19,
  23)\\ &(3, 5, 5, 11, 11) & (3, 6, 7, 14, 15) & (3, 7, 9, 17, 19) &
  (3, 8, 11, 20, 23) & (3, 9, 13, 23, 27)\\ &(3, 6, 6, 15, 15) & (3,
  7, 8, 18, 19) & (3, 8, 10, 21, 23) & (3, 9, 12, 24, 27) & (3, 10,
  14, 27, 31)\\ &(3, 7, 7, 19, 19) & (3, 8, 9, 22, 23) & (3, 9, 11,
  25, 27) & (3, 10, 13, 28, 31) & (3, 11, 15, 31, 35)\\ &(4, 4, 4, 4,
  4) & (4, 5, 6, 7, 8) & (4, 6, 8, 10, 12) & (4, 7, 10, 13, 16) & (4,
  8, 12, 16, 20)\\ &(4, 5, 5, 8, 8) & (4, 6, 7, 11, 12) & (4, 7, 9,
  14, 16) & (4, 8, 11, 17, 20) & (4, 9, 13, 20, 24)\\ &(4, 6, 6, 12,
  12) & (4, 7, 8, 15, 16) & (4, 8, 10, 18, 20) & (4, 9, 12, 21, 24) &
  (4, 10, 14, 24, 28)\\ &(4, 7, 7, 16, 16) & (4, 8, 9, 19, 20) & (4,
  9, 11, 22, 24) & (4, 10, 13, 25, 28) & (4, 11, 15, 28, 32)\\ &(4, 8,
  8, 20, 20) & (4, 9, 10, 23, 24) & (4, 10, 12, 26, 28) & (4, 11, 14,
  29, 32) & (4, 12, 16, 32, 36)\\
\end{supertabular}
\end{center}

\newpage

%\topcaption{}
\tablefirsthead{\hline \multicolumn{6}{|c|}{$G_5$ }\\ \hline}
\tablelasttail{\hline}
%%\topcaption{The elements of $G_5$}
\begin{center}
\begin{supertabular}{|llllll|}
  &(0, 0, 0, 0, 0) & (0, 1, 1, 0, 3) & (0, 2, 2, 0, 6) & (0, 3, 3, 0,
  9) & (0, 4, 4, 0, 12)\\ &(0, 1, 3, 1, 0) & (0, 2, 4, 1, 3) & (0, 3,
  5, 1, 6) & (0, 4, 6, 1, 9) & (0, 5, 7, 1, 12)\\ &(0, 2, 6, 2, 0) &
  (0, 3, 7, 2, 3) & (0, 4, 8, 2, 6) & (0, 5, 9, 2, 9) & (0, 6, 10, 2,
  12)\\ &(0, 3, 9, 3, 0) & (0, 4, 10, 3, 3) & (0, 5, 11, 3, 6) & (0,
  6, 12, 3, 9) & (0, 7, 13, 3, 12)\\ &(0, 4, 12, 4, 0) & (0, 5, 13, 4,
  3) & (0, 6, 14, 4, 6) & (0, 7, 15, 4, 9) & (0, 8, 16, 4, 12)\\ &(1,
  1, 1, 1, 1) & (1, 2, 2, 1, 4) & (1, 3, 3, 1, 7) & (1, 4, 4, 1, 10) &
  (1, 5, 5, 1, 13)\\ &(1, 2, 4, 2, 1) & (1, 3, 5, 2, 4) & (1, 4, 6, 2,
  7) & (1, 5, 7, 2, 10) & (1, 6, 8, 2, 13)\\ &(1, 3, 7, 3, 1) & (1, 4,
  8, 3, 4) & (1, 5, 9, 3, 7) & (1, 6, 10, 3, 10) & (1, 7, 11, 3, 13)\\
  &(1, 4, 10, 4, 1) & (1, 5, 11, 4, 4) & (1, 6, 12, 4, 7) & (1, 7, 13,
  4, 10) & (1, 8, 14, 4, 13)\\ &(1, 5, 13, 5, 1) & (1, 6, 14, 5, 4) &
  (1, 7, 15, 5, 7) & (1, 8, 16, 5, 10) & (1, 9, 17, 5, 13)\\ &(2, 2,
  2, 2, 2) & (2, 3, 3, 2, 5) & (2, 4, 4, 2, 8) & (2, 5, 5, 2, 11) &
  (2, 6, 6, 2, 14)\\ &(2, 3, 5, 3, 2) & (2, 4, 6, 3, 5) & (2, 5, 7, 3,
  8) & (2, 6, 8, 3, 11) & (2, 7, 9, 3, 14)\\ &(2, 4, 8, 4, 2) & (2, 5,
  9, 4, 5) & (2, 6, 10, 4, 8) & (2, 7, 11, 4, 11) & (2, 8, 12, 4,
  14)\\ &(2, 5, 11, 5, 2) & (2, 6, 12, 5, 5) & (2, 7, 13, 5, 8) & (2,
  8, 14, 5, 11) & (2, 9, 15, 5, 14)\\ &(2, 6, 14, 6, 2) & (2, 7, 15,
  6, 5) & (2, 8, 16, 6, 8) & (2, 9, 17, 6, 11) & (2, 10, 18, 6, 14)\\
  &(3, 3, 3, 3, 3) & (3, 4, 4, 3, 6) & (3, 5, 5, 3, 9) & (3, 6, 6, 3,
  12) & (3, 7, 7, 3, 15)\\ &(3, 4, 6, 4, 3) & (3, 5, 7, 4, 6) & (3, 6,
  8, 4, 9) & (3, 7, 9, 4, 12) & (3, 8, 10, 4, 15)\\ &(3, 5, 9, 5, 3) &
  (3, 6, 10, 5, 6) & (3, 7, 11, 5, 9) & (3, 8, 12, 5, 12) & (3, 9, 13,
  5, 15)\\ &(3, 6, 12, 6, 3) & (3, 7, 13, 6, 6) & (3, 8, 14, 6, 9) &
  (3, 9, 15, 6, 12) & (3, 10, 16, 6, 15)\\ &(3, 7, 15, 7, 3) & (3, 8,
  16, 7, 6) & (3, 9, 17, 7, 9) & (3, 10, 18, 7, 12) & (3, 11, 19, 7,
  15)\\ &(4, 4, 4, 4, 4) & (4, 5, 5, 4, 7) & (4, 6, 6, 4, 10) & (4, 7,
  7, 4, 13) & (4, 8, 8, 4, 16)\\ &(4, 5, 7, 5, 4) & (4, 6, 8, 5, 7) &
  (4, 7, 9, 5, 10) & (4, 8, 10, 5, 13) & (4, 9, 11, 5, 16)\\ &(4, 6,
  10, 6, 4) & (4, 7, 11, 6, 7) & (4, 8, 12, 6, 10) & (4, 9, 13, 6, 13)
  & (4, 10, 14, 6, 16)\\ &(4, 7, 13, 7, 4) & (4, 8, 14, 7, 7) & (4, 9,
  15, 7, 10) & (4, 10, 16, 7, 13) & (4, 11, 17, 7, 16)\\ &(4, 8, 16,
  8, 4) & (4, 9, 17, 8, 7) & (4, 10, 18, 8, 10) & (4, 11, 19, 8, 13) &
  (4, 12, 20, 8, 16)\\
\end{supertabular}
\end{center}

\newpage

%\topcaption{}
\tablefirsthead{\hline \multicolumn{6}{|c|}{$G_6$ }\\ \hline}
\tablelasttail{\hline}
%%\topcaption{The elements of $G_6$}
\begin{center}
\begin{supertabular}{|llllll|}
  &(0, 0, 0, 0, 0) & (0, 3, 0, 1, 1) & (0, 6, 0, 2, 2) & (0, 9, 0, 3,
  3) & (0, 12, 0, 4, 4)\\ &(0, 1, 4, 0, 0) & (0, 4, 4, 1, 1) & (0, 7,
  4, 2, 2) & (0, 10, 4, 3, 3) & (0, 13, 4, 4, 4)\\ &(0, 2, 8, 0, 0) &
  (0, 5, 8, 1, 1) & (0, 8, 8, 2, 2) & (0, 11, 8, 3, 3) & (0, 14, 8, 4,
  4)\\ &(0, 3, 12, 0, 0) & (0, 6, 12, 1, 1) & (0, 9, 12, 2, 2) & (0,
  12, 12, 3, 3) & (0, 15, 12, 4, 4)\\ &(0, 4, 16, 0, 0) & (0, 7, 16,
  1, 1) & (0, 10, 16, 2, 2) & (0, 13, 16, 3, 3) & (0, 16, 16, 4, 4)\\
  &(1, 1, 1, 1, 1) & (1, 4, 1, 2, 2) & (1, 7, 1, 3, 3) & (1, 10, 1, 4,
  4) & (1, 13, 1, 5, 5)\\ &(1, 2, 5, 1, 1) & (1, 5, 5, 2, 2) & (1, 8,
  5, 3, 3) & (1, 11, 5, 4, 4) & (1, 14, 5, 5, 5)\\ &(1, 3, 9, 1, 1) &
  (1, 6, 9, 2, 2) & (1, 9, 9, 3, 3) & (1, 12, 9, 4, 4) & (1, 15, 9, 5,
  5)\\ &(1, 4, 13, 1, 1) & (1, 7, 13, 2, 2) & (1, 10, 13, 3, 3) & (1,
  13, 13, 4, 4) & (1, 16, 13, 5, 5)\\ &(1, 5, 17, 1, 1) & (1, 8, 17,
  2, 2) & (1, 11, 17, 3, 3) & (1, 14, 17, 4, 4) & (1, 17, 17, 5, 5)\\
  &(2, 2, 2, 2, 2) & (2, 5, 2, 3, 3) & (2, 8, 2, 4, 4) & (2, 11, 2, 5,
  5) & (2, 14, 2, 6, 6)\\ &(2, 3, 6, 2, 2) & (2, 6, 6, 3, 3) & (2, 9,
  6, 4, 4) & (2, 12, 6, 5, 5) & (2, 15, 6, 6, 6)\\ &(2, 4, 10, 2, 2) &
  (2, 7, 10, 3, 3) & (2, 10, 10, 4, 4) & (2, 13, 10, 5, 5) & (2, 16,
  10, 6, 6)\\ &(2, 5, 14, 2, 2) & (2, 8, 14, 3, 3) & (2, 11, 14, 4, 4)
  & (2, 14, 14, 5, 5) & (2, 17, 14, 6, 6)\\ &(2, 6, 18, 2, 2) & (2, 9,
  18, 3, 3) & (2, 12, 18, 4, 4) & (2, 15, 18, 5, 5) & (2, 18, 18, 6,
  6)\\ &(3, 3, 3, 3, 3) & (3, 6, 3, 4, 4) & (3, 9, 3, 5, 5) & (3, 12,
  3, 6, 6) & (3, 15, 3, 7, 7)\\ &(3, 4, 7, 3, 3) & (3, 7, 7, 4, 4) &
  (3, 10, 7, 5, 5) & (3, 13, 7, 6, 6) & (3, 16, 7, 7, 7)\\ &(3, 5, 11,
  3, 3) & (3, 8, 11, 4, 4) & (3, 11, 11, 5, 5) & (3, 14, 11, 6, 6) &
  (3, 17, 11, 7, 7)\\ &(3, 6, 15, 3, 3) & (3, 9, 15, 4, 4) & (3, 12,
  15, 5, 5) & (3, 15, 15, 6, 6) & (3, 18, 15, 7, 7)\\ &(3, 7, 19, 3,
  3) & (3, 10, 19, 4, 4) & (3, 13, 19, 5, 5) & (3, 16, 19, 6, 6) & (3,
  19, 19, 7, 7)\\ &(4, 4, 4, 4, 4) & (4, 7, 4, 5, 5) & (4, 10, 4, 6,
  6) & (4, 13, 4, 7, 7) & (4, 16, 4, 8, 8)\\ &(4, 5, 8, 4, 4) & (4, 8,
  8, 5, 5) & (4, 11, 8, 6, 6) & (4, 14, 8, 7, 7) & (4, 17, 8, 8, 8)\\
  &(4, 6, 12, 4, 4) & (4, 9, 12, 5, 5) & (4, 12, 12, 6, 6) & (4, 15,
  12, 7, 7) & (4, 18, 12, 8, 8)\\ &(4, 7, 16, 4, 4) & (4, 10, 16, 5,
  5) & (4, 13, 16, 6, 6) & (4, 16, 16, 7, 7) & (4, 19, 16, 8, 8)\\
  &(4, 8, 20, 4, 4) & (4, 11, 20, 5, 5) & (4, 14, 20, 6, 6) & (4, 17,
  20, 7, 7) & (4, 20, 20, 8, 8)\\
\end{supertabular}
\end{center}

\newpage

%\topcaption{}
\tablefirsthead{\hline \multicolumn{6}{|c|}{$G_7$ }\\ \hline}
\tablehead{\multicolumn{6}{l}{{\it cont. from previous page }}\\
  \hline}
\tabletail{\hline \multicolumn{6}{r}{{\it cont. to next page }}\\ }
\tablelasttail{\hline\multicolumn{6}{r}{{ \it end of table }}\\}
%%\topcaption{The elements of $G_7$}
\begin{center}
\begin{supertabular}{|llllll|}
  &(0, 0, 0, 0, 0) & (0, 0, 0, 1, 4) & (0, 0, 0, 2, 8) & (0, 0, 0, 3,
  12) & (0, 0, 0, 4, 16)\\ &(0, 1, 1, 4, 4) & (0, 1, 1, 5, 8) & (0, 1,
  1, 6, 12) & (0, 1, 1, 7, 16) & (0, 1, 1, 8, 20)\\ &(0, 2, 2, 8, 8) &
  (0, 2, 2, 9, 12) & (0, 2, 2, 10, 16) & (0, 2, 2, 11, 20) & (0, 2, 2,
  12, 24)\\ &(0, 3, 3, 12, 12) & (0, 3, 3, 13, 16) & (0, 3, 3, 14, 20)
  & (0, 3, 3, 15, 24) & (0, 3, 3, 16, 28)\\ &(0, 4, 4, 16, 16) & (0,
  4, 4, 17, 20) & (0, 4, 4, 18, 24) & (0, 4, 4, 19, 28) & (0, 4, 4,
  20, 32)\\ &(0, 1, 2, 3, 4) & (0, 1, 2, 4, 8) & (0, 1, 2, 5, 12) &
  (0, 1, 2, 6, 16) & (0, 1, 2, 7, 20)\\ &(0, 2, 3, 7, 8) & (0, 2, 3,
  8, 12) & (0, 2, 3, 9, 16) & (0, 2, 3, 10, 20) & (0, 2, 3, 11, 24)\\
  &(0, 3, 4, 11, 12) & (0, 3, 4, 12, 16) & (0, 3, 4, 13, 20) & (0, 3,
  4, 14, 24) & (0, 3, 4, 15, 28)\\ &(0, 4, 5, 15, 16) & (0, 4, 5, 16,
  20) & (0, 4, 5, 17, 24) & (0, 4, 5, 18, 28) & (0, 4, 5, 19, 32)\\
  &(0, 5, 6, 19, 20) & (0, 5, 6, 20, 24) & (0, 5, 6, 21, 28) & (0, 5,
  6, 22, 32) & (0, 5, 6, 23, 36)\\ &(0, 2, 4, 6, 8) & (0, 2, 4, 7, 12)
  & (0, 2, 4, 8, 16) & (0, 2, 4, 9, 20) & (0, 2, 4, 10, 24)\\ &(0, 3,
  5, 10, 12) & (0, 3, 5, 11, 16) & (0, 3, 5, 12, 20) & (0, 3, 5, 13,
  24) & (0, 3, 5, 14, 28)\\ &(0, 4, 6, 14, 16) & (0, 4, 6, 15, 20) &
  (0, 4, 6, 16, 24) & (0, 4, 6, 17, 28) & (0, 4, 6, 18, 32)\\ &(0, 5,
  7, 18, 20) & (0, 5, 7, 19, 24) & (0, 5, 7, 20, 28) & (0, 5, 7, 21,
  32) & (0, 5, 7, 22, 36)\\ &(0, 6, 8, 22, 24) & (0, 6, 8, 23, 28) &
  (0, 6, 8, 24, 32) & (0, 6, 8, 25, 36) & (0, 6, 8, 26, 40)\\ &(0, 3,
  6, 9, 12) & (0, 3, 6, 10, 16) & (0, 3, 6, 11, 20) & (0, 3, 6, 12,
  24) & (0, 3, 6, 13, 28)\\ &(0, 4, 7, 13, 16) & (0, 4, 7, 14, 20) &
  (0, 4, 7, 15, 24) & (0, 4, 7, 16, 28) & (0, 4, 7, 17, 32)\\ &(0, 5,
  8, 17, 20) & (0, 5, 8, 18, 24) & (0, 5, 8, 19, 28) & (0, 5, 8, 20,
  32) & (0, 5, 8, 21, 36)\\ &(0, 6, 9, 21, 24) & (0, 6, 9, 22, 28) &
  (0, 6, 9, 23, 32) & (0, 6, 9, 24, 36) & (0, 6, 9, 25, 40)\\ &(0, 7,
  10, 25, 28) & (0, 7, 10, 26, 32) & (0, 7, 10, 27, 36) & (0, 7, 10,
  28, 40) & (0, 7, 10, 29, 44)\\ &(0, 4, 8, 12, 16) & (0, 4, 8, 13,
  20) & (0, 4, 8, 14, 24) & (0, 4, 8, 15, 28) & (0, 4, 8, 16, 32)\\
  &(0, 5, 9, 16, 20) & (0, 5, 9, 17, 24) & (0, 5, 9, 18, 28) & (0, 5,
  9, 19, 32) & (0, 5, 9, 20, 36)\\ &(0, 6, 10, 20, 24) & (0, 6, 10,
  21, 28) & (0, 6, 10, 22, 32) & (0, 6, 10, 23, 36) & (0, 6, 10, 24,
  40)\\ &(0, 7, 11, 24, 28) & (0, 7, 11, 25, 32) & (0, 7, 11, 26, 36)
  & (0, 7, 11, 27, 40) & (0, 7, 11, 28, 44)\\ &(0, 8, 12, 28, 32) &
  (0, 8, 12, 29, 36) & (0, 8, 12, 30, 40) & (0, 8, 12, 31, 44) & (0,
  8, 12, 32, 48)\\ &(1, 1, 1, 1, 1) & (1, 1, 1, 2, 5) & (1, 1, 1, 3,
  9) & (1, 1, 1, 4, 13) & (1, 1, 1, 5, 17)\\ &(1, 2, 2, 5, 5) & (1, 2,
  2, 6, 9) & (1, 2, 2, 7, 13) & (1, 2, 2, 8, 17) & (1, 2, 2, 9, 21)\\
  &(1, 3, 3, 9, 9) & (1, 3, 3, 10, 13) & (1, 3, 3, 11, 17) & (1, 3, 3,
  12, 21) & (1, 3, 3, 13, 25)\\ &(1, 4, 4, 13, 13) & (1, 4, 4, 14, 17)
  & (1, 4, 4, 15, 21) & (1, 4, 4, 16, 25) & (1, 4, 4, 17, 29)\\ &(1,
  5, 5, 17, 17) & (1, 5, 5, 18, 21) & (1, 5, 5, 19, 25) & (1, 5, 5,
  20, 29) & (1, 5, 5, 21, 33)\\ &(1, 2, 3, 4, 5) & (1, 2, 3, 5, 9) &
  (1, 2, 3, 6, 13) & (1, 2, 3, 7, 17) & (1, 2, 3, 8, 21)\\ &(1, 3, 4,
  8, 9) & (1, 3, 4, 9, 13) & (1, 3, 4, 10, 17) & (1, 3, 4, 11, 21) &
  (1, 3, 4, 12, 25)\\ &(1, 4, 5, 12, 13) & (1, 4, 5, 13, 17) & (1, 4,
  5, 14, 21) & (1, 4, 5, 15, 25) & (1, 4, 5, 16, 29)\\ &(1, 5, 6, 16,
  17) & (1, 5, 6, 17, 21) & (1, 5, 6, 18, 25) & (1, 5, 6, 19, 29) &
  (1, 5, 6, 20, 33)\\ &(1, 6, 7, 20, 21) & (1, 6, 7, 21, 25) & (1, 6,
  7, 22, 29) & (1, 6, 7, 23, 33) & (1, 6, 7, 24, 37)\\ &(1, 3, 5, 7,
  9) & (1, 3, 5, 8, 13) & (1, 3, 5, 9, 17) & (1, 3, 5, 10, 21) & (1,
  3, 5, 11, 25)\\ &(1, 4, 6, 11, 13) & (1, 4, 6, 12, 17) & (1, 4, 6,
  13, 21) & (1, 4, 6, 14, 25) & (1, 4, 6, 15, 29)\\ &(1, 5, 7, 15, 17)
  & (1, 5, 7, 16, 21) & (1, 5, 7, 17, 25) & (1, 5, 7, 18, 29) & (1, 5,
  7, 19, 33)\\ &(1, 6, 8, 19, 21) & (1, 6, 8, 20, 25) & (1, 6, 8, 21,
  29) & (1, 6, 8, 22, 33) & (1, 6, 8, 23, 37)\\ &(1, 7, 9, 23, 25) &
  (1, 7, 9, 24, 29) & (1, 7, 9, 25, 33) & (1, 7, 9, 26, 37) & (1, 7,
  9, 27, 41)\\ &(1, 4, 7, 10, 13) & (1, 4, 7, 11, 17) & (1, 4, 7, 12,
  21) & (1, 4, 7, 13, 25) & (1, 4, 7, 14, 29)\\ &(1, 5, 8, 14, 17) &
  (1, 5, 8, 15, 21) & (1, 5, 8, 16, 25) & (1, 5, 8, 17, 29) & (1, 5,
  8, 18, 33)\\ &(1, 6, 9, 18, 21) & (1, 6, 9, 19, 25) & (1, 6, 9, 20,
  29) & (1, 6, 9, 21, 33) & (1, 6, 9, 22, 37)\\ &(1, 7, 10, 22, 25) &
  (1, 7, 10, 23, 29) & (1, 7, 10, 24, 33) & (1, 7, 10, 25, 37) & (1,
  7, 10, 26, 41)\\ &(1, 8, 11, 26, 29) & (1, 8, 11, 27, 33) & (1, 8,
  11, 28, 37) & (1, 8, 11, 29, 41) & (1, 8, 11, 30, 45)\\ &(1, 5, 9,
  13, 17) & (1, 5, 9, 14, 21) & (1, 5, 9, 15, 25) & (1, 5, 9, 16, 29)
  & (1, 5, 9, 17, 33)\\ &(1, 6, 10, 17, 21) & (1, 6, 10, 18, 25) & (1,
  6, 10, 19, 29) & (1, 6, 10, 20, 33) & (1, 6, 10, 21, 37)\\ &(1, 7,
  11, 21, 25) & (1, 7, 11, 22, 29) & (1, 7, 11, 23, 33) & (1, 7, 11,
  24, 37) & (1, 7, 11, 25, 41)\\ &(1, 8, 12, 25, 29) & (1, 8, 12, 26,
  33) & (1, 8, 12, 27, 37) & (1, 8, 12, 28, 41) & (1, 8, 12, 29, 45)\\
  &(1, 9, 13, 29, 33) & (1, 9, 13, 30, 37) & (1, 9, 13, 31, 41) & (1,
  9, 13, 32, 45) & (1, 9, 13, 33, 49)\\ &(2, 2, 2, 2, 2) & (2, 2, 2,
  3, 6) & (2, 2, 2, 4, 10) & (2, 2, 2, 5, 14) & (2, 2, 2, 6, 18)\\
  &(2, 3, 3, 6, 6) & (2, 3, 3, 7, 10) & (2, 3, 3, 8, 14) & (2, 3, 3,
  9, 18) & (2, 3, 3, 10, 22)\\ &(2, 4, 4, 10, 10) & (2, 4, 4, 11, 14)
  & (2, 4, 4, 12, 18) & (2, 4, 4, 13, 22) & (2, 4, 4, 14, 26)\\ &(2,
  5, 5, 14, 14) & (2, 5, 5, 15, 18) & (2, 5, 5, 16, 22) & (2, 5, 5,
  17, 26) & (2, 5, 5, 18, 30)\\ &(2, 6, 6, 18, 18) & (2, 6, 6, 19, 22)
  & (2, 6, 6, 20, 26) & (2, 6, 6, 21, 30) & (2, 6, 6, 22, 34)\\ &(2,
  3, 4, 5, 6) & (2, 3, 4, 6, 10) & (2, 3, 4, 7, 14) & (2, 3, 4, 8, 18)
  & (2, 3, 4, 9, 22)\\ &(2, 4, 5, 9, 10) & (2, 4, 5, 10, 14) & (2, 4,
  5, 11, 18) & (2, 4, 5, 12, 22) & (2, 4, 5, 13, 26)\\ &(2, 5, 6, 13,
  14) & (2, 5, 6, 14, 18) & (2, 5, 6, 15, 22) & (2, 5, 6, 16, 26) &
  (2, 5, 6, 17, 30)\\ &(2, 6, 7, 17, 18) & (2, 6, 7, 18, 22) & (2, 6,
  7, 19, 26) & (2, 6, 7, 20, 30) & (2, 6, 7, 21, 34)\\ &(2, 7, 8, 21,
  22) & (2, 7, 8, 22, 26) & (2, 7, 8, 23, 30) & (2, 7, 8, 24, 34) &
  (2, 7, 8, 25, 38)\\ &(2, 4, 6, 8, 10) & (2, 4, 6, 9, 14) & (2, 4, 6,
  10, 18) & (2, 4, 6, 11, 22) & (2, 4, 6, 12, 26)\\ &(2, 5, 7, 12, 14)
  & (2, 5, 7, 13, 18) & (2, 5, 7, 14, 22) & (2, 5, 7, 15, 26) & (2, 5,
  7, 16, 30)\\ &(2, 6, 8, 16, 18) & (2, 6, 8, 17, 22) & (2, 6, 8, 18,
  26) & (2, 6, 8, 19, 30) & (2, 6, 8, 20, 34)\\ &(2, 7, 9, 20, 22) &
  (2, 7, 9, 21, 26) & (2, 7, 9, 22, 30) & (2, 7, 9, 23, 34) & (2, 7,
  9, 24, 38)\\ &(2, 8, 10, 24, 26) & (2, 8, 10, 25, 30) & (2, 8, 10,
  26, 34) & (2, 8, 10, 27, 38) & (2, 8, 10, 28, 42)\\ &(2, 5, 8, 11,
  14) & (2, 5, 8, 12, 18) & (2, 5, 8, 13, 22) & (2, 5, 8, 14, 26) &
  (2, 5, 8, 15, 30)\\ &(2, 6, 9, 15, 18) & (2, 6, 9, 16, 22) & (2, 6,
  9, 17, 26) & (2, 6, 9, 18, 30) & (2, 6, 9, 19, 34)\\ &(2, 7, 10, 19,
  22) & (2, 7, 10, 20, 26) & (2, 7, 10, 21, 30) & (2, 7, 10, 22, 34) &
  (2, 7, 10, 23, 38)\\ &(2, 8, 11, 23, 26) & (2, 8, 11, 24, 30) & (2,
  8, 11, 25, 34) & (2, 8, 11, 26, 38) & (2, 8, 11, 27, 42)\\ &(2, 9,
  12, 27, 30) & (2, 9, 12, 28, 34) & (2, 9, 12, 29, 38) & (2, 9, 12,
  30, 42) & (2, 9, 12, 31, 46)\\ &(2, 6, 10, 14, 18) & (2, 6, 10, 15,
  22) & (2, 6, 10, 16, 26) & (2, 6, 10, 17, 30) & (2, 6, 10, 18, 34)\\
  &(2, 7, 11, 18, 22) & (2, 7, 11, 19, 26) & (2, 7, 11, 20, 30) & (2,
  7, 11, 21, 34) & (2, 7, 11, 22, 38)\\ &(2, 8, 12, 22, 26) & (2, 8,
  12, 23, 30) & (2, 8, 12, 24, 34) & (2, 8, 12, 25, 38) & (2, 8, 12,
  26, 42)\\ &(2, 9, 13, 26, 30) & (2, 9, 13, 27, 34) & (2, 9, 13, 28,
  38) & (2, 9, 13, 29, 42) & (2, 9, 13, 30, 46)\\ &(2, 10, 14, 30, 34)
  & (2, 10, 14, 31, 38) & (2, 10, 14, 32, 42) & (2, 10, 14, 33, 46) &
  (2, 10, 14, 34, 50)\\ &(3, 3, 3, 3, 3) & (3, 3, 3, 4, 7) & (3, 3, 3,
  5, 11) & (3, 3, 3, 6, 15) & (3, 3, 3, 7, 19)\\ &(3, 4, 4, 7, 7) &
  (3, 4, 4, 8, 11) & (3, 4, 4, 9, 15) & (3, 4, 4, 10, 19) & (3, 4, 4,
  11, 23)\\ &(3, 5, 5, 11, 11) & (3, 5, 5, 12, 15) & (3, 5, 5, 13, 19)
  & (3, 5, 5, 14, 23) & (3, 5, 5, 15, 27)\\ &(3, 6, 6, 15, 15) & (3,
  6, 6, 16, 19) & (3, 6, 6, 17, 23) & (3, 6, 6, 18, 27) & (3, 6, 6,
  19, 31)\\ &(3, 7, 7, 19, 19) & (3, 7, 7, 20, 23) & (3, 7, 7, 21, 27)
  & (3, 7, 7, 22, 31) & (3, 7, 7, 23, 35)\\ &(3, 4, 5, 6, 7) & (3, 4,
  5, 7, 11) & (3, 4, 5, 8, 15) & (3, 4, 5, 9, 19) & (3, 4, 5, 10,
  23)\\ &(3, 5, 6, 10, 11) & (3, 5, 6, 11, 15) & (3, 5, 6, 12, 19) &
  (3, 5, 6, 13, 23) & (3, 5, 6, 14, 27)\\ &(3, 6, 7, 14, 15) & (3, 6,
  7, 15, 19) & (3, 6, 7, 16, 23) & (3, 6, 7, 17, 27) & (3, 6, 7, 18,
  31)\\ &(3, 7, 8, 18, 19) & (3, 7, 8, 19, 23) & (3, 7, 8, 20, 27) &
  (3, 7, 8, 21, 31) & (3, 7, 8, 22, 35)\\ &(3, 8, 9, 22, 23) & (3, 8,
  9, 23, 27) & (3, 8, 9, 24, 31) & (3, 8, 9, 25, 35) & (3, 8, 9, 26,
  39)\\ &(3, 5, 7, 9, 11) & (3, 5, 7, 10, 15) & (3, 5, 7, 11, 19) &
  (3, 5, 7, 12, 23) & (3, 5, 7, 13, 27)\\ &(3, 6, 8, 13, 15) & (3, 6,
  8, 14, 19) & (3, 6, 8, 15, 23) & (3, 6, 8, 16, 27) & (3, 6, 8, 17,
  31)\\ &(3, 7, 9, 17, 19) & (3, 7, 9, 18, 23) & (3, 7, 9, 19, 27) &
  (3, 7, 9, 20, 31) & (3, 7, 9, 21, 35)\\ &(3, 8, 10, 21, 23) & (3, 8,
  10, 22, 27) & (3, 8, 10, 23, 31) & (3, 8, 10, 24, 35) & (3, 8, 10,
  25, 39)\\ &(3, 9, 11, 25, 27) & (3, 9, 11, 26, 31) & (3, 9, 11, 27,
  35) & (3, 9, 11, 28, 39) & (3, 9, 11, 29, 43)\\ &(3, 6, 9, 12, 15) &
  (3, 6, 9, 13, 19) & (3, 6, 9, 14, 23) & (3, 6, 9, 15, 27) & (3, 6,
  9, 16, 31)\\ &(3, 7, 10, 16, 19) & (3, 7, 10, 17, 23) & (3, 7, 10,
  18, 27) & (3, 7, 10, 19, 31) & (3, 7, 10, 20, 35)\\ &(3, 8, 11, 20,
  23) & (3, 8, 11, 21, 27) & (3, 8, 11, 22, 31) & (3, 8, 11, 23, 35) &
  (3, 8, 11, 24, 39)\\ &(3, 9, 12, 24, 27) & (3, 9, 12, 25, 31) & (3,
  9, 12, 26, 35) & (3, 9, 12, 27, 39) & (3, 9, 12, 28, 43)\\ &(3, 10,
  13, 28, 31) & (3, 10, 13, 29, 35) & (3, 10, 13, 30, 39) & (3, 10,
  13, 31, 43) & (3, 10, 13, 32, 47)\\ &(3, 7, 11, 15, 19) & (3, 7, 11,
  16, 23) & (3, 7, 11, 17, 27) & (3, 7, 11, 18, 31) & (3, 7, 11, 19,
  35)\\ &(3, 8, 12, 19, 23) & (3, 8, 12, 20, 27) & (3, 8, 12, 21, 31)
  & (3, 8, 12, 22, 35) & (3, 8, 12, 23, 39)\\ &(3, 9, 13, 23, 27) &
  (3, 9, 13, 24, 31) & (3, 9, 13, 25, 35) & (3, 9, 13, 26, 39) & (3,
  9, 13, 27, 43)\\ &(3, 10, 14, 27, 31) & (3, 10, 14, 28, 35) & (3,
  10, 14, 29, 39) & (3, 10, 14, 30, 43) & (3, 10, 14, 31, 47)\\ &(3,
  11, 15, 31, 35) & (3, 11, 15, 32, 39) & (3, 11, 15, 33, 43) & (3,
  11, 15, 34, 47) & (3, 11, 15, 35, 51)\\ &(4, 4, 4, 4, 4) & (4, 4, 4,
  5, 8) & (4, 4, 4, 6, 12) & (4, 4, 4, 7, 16) & (4, 4, 4, 8, 20)\\
  &(4, 5, 5, 8, 8) & (4, 5, 5, 9, 12) & (4, 5, 5, 10, 16) & (4, 5, 5,
  11, 20) & (4, 5, 5, 12, 24)\\ &(4, 6, 6, 12, 12) & (4, 6, 6, 13, 16)
  & (4, 6, 6, 14, 20) & (4, 6, 6, 15, 24) & (4, 6, 6, 16, 28)\\ &(4,
  7, 7, 16, 16) & (4, 7, 7, 17, 20) & (4, 7, 7, 18, 24) & (4, 7, 7,
  19, 28) & (4, 7, 7, 20, 32)\\ &(4, 8, 8, 20, 20) & (4, 8, 8, 21, 24)
  & (4, 8, 8, 22, 28) & (4, 8, 8, 23, 32) & (4, 8, 8, 24, 36)\\ &(4,
  5, 6, 7, 8) & (4, 5, 6, 8, 12) & (4, 5, 6, 9, 16) & (4, 5, 6, 10,
  20) & (4, 5, 6, 11, 24)\\ &(4, 6, 7, 11, 12) & (4, 6, 7, 12, 16) &
  (4, 6, 7, 13, 20) & (4, 6, 7, 14, 24) & (4, 6, 7, 15, 28)\\ &(4, 7,
  8, 15, 16) & (4, 7, 8, 16, 20) & (4, 7, 8, 17, 24) & (4, 7, 8, 18,
  28) & (4, 7, 8, 19, 32)\\ &(4, 8, 9, 19, 20) & (4, 8, 9, 20, 24) &
  (4, 8, 9, 21, 28) & (4, 8, 9, 22, 32) & (4, 8, 9, 23, 36)\\ &(4, 9,
  10, 23, 24) & (4, 9, 10, 24, 28) & (4, 9, 10, 25, 32) & (4, 9, 10,
  26, 36) & (4, 9, 10, 27, 40)\\ &(4, 6, 8, 10, 12) & (4, 6, 8, 11,
  16) & (4, 6, 8, 12, 20) & (4, 6, 8, 13, 24) & (4, 6, 8, 14, 28)\\
  &(4, 7, 9, 14, 16) & (4, 7, 9, 15, 20) & (4, 7, 9, 16, 24) & (4, 7,
  9, 17, 28) & (4, 7, 9, 18, 32)\\ &(4, 8, 10, 18, 20) & (4, 8, 10,
  19, 24) & (4, 8, 10, 20, 28) & (4, 8, 10, 21, 32) & (4, 8, 10, 22,
  36)\\ &(4, 9, 11, 22, 24) & (4, 9, 11, 23, 28) & (4, 9, 11, 24, 32)
  & (4, 9, 11, 25, 36) & (4, 9, 11, 26, 40)\\ &(4, 10, 12, 26, 28) &
  (4, 10, 12, 27, 32) & (4, 10, 12, 28, 36) & (4, 10, 12, 29, 40) &
  (4, 10, 12, 30, 44)\\ &(4, 7, 10, 13, 16) & (4, 7, 10, 14, 20) & (4,
  7, 10, 15, 24) & (4, 7, 10, 16, 28) & (4, 7, 10, 17, 32)\\ &(4, 8,
  11, 17, 20) & (4, 8, 11, 18, 24) & (4, 8, 11, 19, 28) & (4, 8, 11,
  20, 32) & (4, 8, 11, 21, 36)\\ &(4, 9, 12, 21, 24) & (4, 9, 12, 22,
  28) & (4, 9, 12, 23, 32) & (4, 9, 12, 24, 36) & (4, 9, 12, 25, 40)\\
  &(4, 10, 13, 25, 28) & (4, 10, 13, 26, 32) & (4, 10, 13, 27, 36) &
  (4, 10, 13, 28, 40) & (4, 10, 13, 29, 44)\\ &(4, 11, 14, 29, 32) &
  (4, 11, 14, 30, 36) & (4, 11, 14, 31, 40) & (4, 11, 14, 32, 44) &
  (4, 11, 14, 33, 48)\\ &(4, 8, 12, 16, 20) & (4, 8, 12, 17, 24) & (4,
  8, 12, 18, 28) & (4, 8, 12, 19, 32) & (4, 8, 12, 20, 36)\\ &(4, 9,
  13, 20, 24) & (4, 9, 13, 21, 28) & (4, 9, 13, 22, 32) & (4, 9, 13,
  23, 36) & (4, 9, 13, 24, 40)\\ &(4, 10, 14, 24, 28) & (4, 10, 14,
  25, 32) & (4, 10, 14, 26, 36) & (4, 10, 14, 27, 40) & (4, 10, 14,
  28, 44)\\ &(4, 11, 15, 28, 32) & (4, 11, 15, 29, 36) & (4, 11, 15,
  30, 40) & (4, 11, 15, 31, 44) & (4, 11, 15, 32, 48)\\ &(4, 12, 16,
  32, 36) & (4, 12, 16, 33, 40) & (4, 12, 16, 34, 44) & (4, 12, 16,
  35, 48) & (4, 12, 16, 36, 52)\\
\end{supertabular}
\end{center}

\newpage

%\topcaption{Twisted sector contributions to $\chi_y$ of ${\cal M}_{G_0}$}
\tablefirsthead{\hline \multicolumn{1}{|c|}{$\chi_y^\alpha[W/\!/G_0]$}\\
\hline}
%\tablehead{\hline}
%\tabletail{\hline\\}
\tablelasttail{\hline}
\begin{center}
\begin{supertabular}{|l|}
$ 1 + 101y + 101{y^2} + {y^3},-{y^3},-{y^2},-y,-1$\\
\end{supertabular}
\end{center}
\vspace{2cm}

%\topcaption{Twisted sector contributions to $\chi_y$ of ${\cal M}_{G_1}$}
\tablefirsthead{\hline \multicolumn{1}{|c|}{$\chi_y^\alpha[W/\!/G_1]$}\\
\hline}
%\tablehead{\hline}
%\tabletail{\hline\\}
\tablelasttail{\hline}
\begin{center}
\begin{supertabular}{|l|}
 $1+25y+25{y^2}+{y^3},6y+6{y^2},6y+6{y^2},6y+6{y^2},6y+6{y^2},$\\
  $ -{y^3},0,-{y^2},-{y^2},0,$\\
  $ -{y^2},-{y^2},0,0,-{y^2},$\\
  $ -y,-y,0,0,-y,$\\
 $ -1,0,-y,-y,0 $ \\
\end{supertabular}
\end{center}
\vspace{2cm}

%\topcaption{Twisted sector contributions to $\chi_y$ of ${\cal M}_{G_2}$}
\tablefirsthead{\hline \multicolumn{1}{|c|}{$\chi_y^\alpha[W/\!/G_2]$}\\
\hline}
%\tablehead{\hline}
%\tabletail{\hline\\}
\tablelasttail{\hline}
\begin{center}
\begin{supertabular}{|l|}
  $1+21y+21{y^2}+{y^3},0,0,0,0,$\\
  $ -{y^3},0,0,0,0,$\\
  $ -{y^2},0,0,0,0,$\\
  $ -y,0,0,0,0,$\\
  $ -1,0,0,0,0 $\\
\end{supertabular}
\end{center}
\vspace{2cm}

%\topcaption{Twisted sector contributions to $\chi_y$ of ${\cal M}_{G_3}$}
\tablefirsthead{\hline \multicolumn{1}{|c|}{$\chi_y^\alpha[W/\!/G_3]$}\\
\hline}
%\tablehead{\hline}
%\tabletail{\hline\\}
\tablelasttail{\hline}
\begin{center}
\begin{supertabular}{|l|}
$ 1+17y+17{y^2}+{y^3},0,0,0,0,$\\
  $ -{y^3},-4{y^2},-y,-y,-4{y^2},$\\
  $ -{y^2},-{y^2},-4y,-4y,-{y^2},$\\
  $ -y,-y,-4{y^2},-4{y^2},-y,$\\
  $ -1,-4y,-{y^2},-{y^2},-4y $\\
\end{supertabular}
\end{center}
\newpage

%\topcaption{Twisted sector contributions to $\chi_y$ of ${\cal M}_{G_4}$}
\tablefirsthead{\hline \multicolumn{4}{|c|}{$\chi_y^\alpha[W/\!/G_4]$}\\
\hline}
%\tablehead{\hline}
%\tabletail{\hline\\}
\tablelasttail{\hline}
\begin{center}
\begin{supertabular}{|llll|}
  &$ 1+5y+5{y^2}+{y^3},0,0,0,0,$
  &$ 0,0,2y+2{y^2},0,2y+2{y^2},$
  &$ 0,0,0,2y+2{y^2},2y+2{y^2},$\\
  &$ 0,2y+2{y^2},2y+2{y^2},0,0,$
  &$ 0,2y+2{y^2},0,2y+2{y^2},0,$
  &$ -{y^3},0,0,0,0,$\\
 &$  0,-y,-{y^2},0,0,$
  &$ -y,0,0,-{y^2},0,$
  &$ -y,0,-{y^2},0,0,$\\
 &$  0,0,0,-{y^2},-y,$
 &$  -{y^2},0,0,0,0,$
 &$  -{y^2},0,0,0,-{y^2},$\\
 &$  0,0,-{y^2},0,-{y^2},$
 &$  0,-{y^2},0,-{y^2},0,$
 &$  -{y^2},-{y^2},0,0,0,$\\
 &$  -y,0,0,0,0,$
 &$  -y,0,0,0,-y,$
 &$  0,0,-y,0,-y,$\\
  &$ 0,-y,0,-y,0,$
 &$  -y,-y,0,0,0,$
  &$ -1,0,0,0,0,$\\
 &$  0,-{y^2},-y,0,0,$
 &$  -{y^2},0,0,-y,0,$
 &$  -{y^2},0,-y,0,0,$\\
 &$  0,0,0,-y,-{y^2}$& & \\
\end{supertabular}
\end{center}
\vspace{.5cm}

%\topcaption{Twisted sector contributions to $\chi_y$ of ${\cal M}_{G_5}$}
\tablefirsthead{\hline \multicolumn{4}{|c|}{$\chi_y^\alpha[W/\!/G_5]$ }\\
\hline}
%\tablehead{\hline}
%\tabletail{\hline\\}
\tablelasttail{\hline}
\begin{center}
\begin{supertabular}{|llll|}
&$ 1+y+{y^2}+{y^3},0,0,0,0,$
 &$  0,0,0,0,0,$
 &$  0,0,0,0,0,$\\
 &$  0,0,0,0,0,$
 &$  0,0,0,0,0,$
 &$ -{y^3},-{y^2},-{y^2},0,0,$\\
 &$ -{y^2},0,-{y^2},0,-{y^2},$
 &$ -{y^2},-y,0,0,-{y^2},$
 &$  0,0,-{y^2},0,-y,$\\
  &$ 0,0,0,0,0,$
 &$  -{y^2},0,-y,0,-{y^2},$
  &$ 0,0,0,-{y^2},-y,$\\
 &$  -y,0,0,-{y^2},-y,$
 &$  0,0,0,0,0,$
  &$ -{y^2},0,-{y^2},-{y^2},0,$\\
 &$  -y,-y,0,-{y^2},0,$
  &$ -y,0,-y,-y,0,$
 &$  0,0,0,0,0,$\\
  &$ -{y^2},-{y^2},-y,0,0,$
 &$  0,-{y^2},-y,0,0,$
 &$  -1,0,0,-y,-y,$\\
 &$  0,0,0,0,0,$
 &$  0,-{y^2},0,-y,0,$
 &$  -y,-y,0,0,-{y^2},$\\
 &$  -y,-y,0,-y,0$&  &\\
\end{supertabular}
\end{center}
\vspace{.5cm}

%\topcaption{Twisted sector contributions to $\chi_y$ of ${\cal M}_{G_6}$}
\tablefirsthead{\hline \multicolumn{4}{|c|}{$\chi_y^\alpha[W/\!/G_6]$ }\\
\hline}
%\tablehead{\hline}
%\tabletail{\hline\\}
\tablelasttail{\hline}
\begin{center}
\begin{supertabular}{|llll|}
 &$ 1+5y+5{y^2}+{y^3},0,0,0,0,$
  &$  0,0,0,0,0,$
  &$  0,0,0,0,0,$\\
  &$  0,0,0,0,0,$
   &$ 0,0,0,0,0,$
   &$ -{y^3},-{y^2},-{y^2},0,-4{y^2},$\\
  &$  0,0,0,0,0,$
   &$ -{y^2},-{y^2},-y,-y,0,$
  &$  -{y^2},-{y^2},0,-y,-4{y^2},$\\
  &$  0,-{y^2},-{y^2},-y,-4{y^2},$
 &$  -{y^2},0,-y,-4{y^2},-{y^2},$
 &$  -{y^2},-{y^2},-y,-4{y^2},0,$\\
  &$  0,0,0,0,0,$
   &$ 0,-y,-y,-4y,-{y^2},$
   &$ -{y^2},-y,-y,0,-{y^2},$\\
  &$  -y,-y,-4y,-{y^2},0,$
  &$  -y,-y,0,-{y^2},-{y^2},$
  &$  0,-y,-4{y^2},-{y^2},-{y^2},$\\
  &$  0,0,0,0,0,$
  &$  -y,0,-4y,-{y^2},-y,$
  &$  -1,-4y,0,-y,-y,$\\
  &$  0,-4y,-{y^2},-y,-y,$
  &$  -y,-4y,-{y^2},0,-y,$
  &$  -y,0,-{y^2},-{y^2},-y,$\\
  &$  0,0,0,0,0 $ & &\\
\end{supertabular}
\end{center}

\newpage

%\topcaption{Twisted sector contributions to $\chi_y$ of ${\cal M}_{G_7}$}
\tablefirsthead{\hline \multicolumn{4}{|c|}{$\chi_y^\alpha[W/\!/G_7]$}\\
\hline}
%\tablehead{\hline}
%\tabletail{\hline\\}
\tablelasttail{\hline}
\begin{center}
\begin{supertabular}{|llll|}
  &$ 1 + y + {y^2} + {y^3},0,0,0,0,$
  &$ 0,0,0,0,0,$
  &$ 0,0,0,0,0,$\\
  &$ 0,0,0,0,0,$
  &$ 0,0,0,0,0,$
  &$ 0,0,0,0,0,$\\
  &$ 0,0,0,0,0,$
  &$ 0,0,0,0,0,$
  &$ 0,0,0,0,0,$\\
  &$ 0,0,0,0,0,$
  &$ 0,0,0,0,0,$
  &$ 0,0,0,0,0,$\\
  &$ 0,0,0,0,0,$
  &$ 0,0,0,0,0,$
  &$ 0,0,0,0,0,$\\
  &$ 0,0,0,0,0,$
  &$ 0,0,0,0,0,$
  &$ 0,0,0,0,0,$\\
  &$ 0,0,0,0,0,$
  &$ 0,0,0,0,0,$
  &$ 0,0,0,0,0,$\\
  &$ 0,0,0,0,0,$
  &$ 0,0,0,0,0,$
  &$ 0,0,0,0,0,$\\
  &$ 0,0,0,0,0,$
  &$ -{y^3},0,-{y^2},-{y^2},0,$
  &$ 0,-{y^2},-{y^2},-{y^2},-{y^2},$\\
  &$ -y,0,-{y^2},-{y^2},0,$
  &$ -y,-y,0,0,-y,$
  &$ 0,0,0,0,0,$\\
  &$ 0,0,-{y^2},-{y^2},-{y^2},$
  &$ -y,-y,0,-{y^2},0,$
  &$ 0,0,0,0,0,$\\
  &$ 0,0,0,0,0,$
  &$ 0,0,-{y^2},-{y^2},-{y^2},$
  &$ 0,0,0,0,0,$\\
  &$ -{y^2},-{y^2},-{y^2},0,0,$
  &$ 0,0,0,0,0,$
  &$ -{y^2},0,-{y^2},-{y^2},-{y^2},$\\
  &$ 0,-y,0,-{y^2},-{y^2},$
  &$ 0,-{y^2},-{y^2},0,-y,$
  &$ 0,0,0,0,0,$\\
  &$ -{y^2},0,0,-{y^2},-{y^2},$
  &$ 0,0,0,0,0,$
  &$ -{y^2},-{y^2},-{y^2},-{y^2},0,$\\
  &$ 0,0,0,0,0,$
  &$ 0,0,0,0,0,$
  &$ 0,-{y^2},-{y^2},-{y^2},0,$\\
  &$ 0,-{y^2},-{y^2},-{y^2},0,$
  &$ -y,0,-{y^2},0,-y,$
  &$ -{y^2},-{y^2},0,0,-{y^2},$\\
  &$ -{y^2},0,-y,-y,0,$
  &$ 0,-y,-y,-y,-y,$
  &$ 0,0,0,0,0,$\\
  &$ -{y^2},-{y^2},0,0,-{y^2},$
  &$ 0,0,-y,-y,-y,$
  &$ 0,0,0,0,0,$\\
  &$ 0,0,0,0,0,$
  &$ -{y^2},-{y^2},-{y^2},0,-{y^2},$
  &$ -{y^2},-{y^2},0,-y,0,$\\
  &$ 0,-y,0,-{y^2},-{y^2},$
  &$ 0,0,0,0,0,$
  &$ -{y^2},-{y^2},-{y^2},0,0,$\\
  &$ 0,-{y^2},0,-y,-y,$
  &$ 0,0,0,0,0,$
  &$ 0,0,0,0,0,$\\
  &$ 0,-{y^2},-{y^2},0,-y,$
  &$ 0,0,0,0,0,$
 &$ -{y^2},0,0,-{y^2},-{y^2},$\\
  &$ 0,-y,-y,0,-{y^2},$
  &$ 0,0,0,0,0,$
 &$  -{y^2},-{y^2},0,-{y^2},-{y^2},$\\
  &$ -{y^2},0,-y,0,-{y^2},$
  &$ 0,-y,-y,-y,0,$
  &$ 0,0,0,0,0,$\\
  &$ -y,-y,0,0,-y,$
  &$ -y,-y,0,0,-y,$
  &$ 0,0,0,0,0,$\\
  &$ 0,-{y^2},-{y^2},-{y^2},-{y^2},$
  &$ -y,0,-{y^2},-{y^2},0,$
  &$ 0,0,0,0,0,$\\
  &$ 0,0,0,0,0,$
  &$ 0,0,-{y^2},-{y^2},-{y^2},$
  &$ -y,-y,0,-{y^2},0,$\\
  &$ -y,-y,-y,0,-y,$
  &$ 0,0,0,0,0,$
  &$ 0,-y,0,-{y^2},-{y^2},$\\
 &$  -y,-y,-y,0,0,$
  &$ 0,0,0,0,0,$
 &$  0,-{y^2},0,-y,-y,$\\
  &$ 0,-y,-y,0,-{y^2},$
  &$ 0,0,0,0,0,$
  &$ 0,-{y^2},-{y^2},0,-y,$\\
  &$ -y,0,0,-y,-y,$
 &$  0,0,0,0,0,$
  &$ 0,-{y^2},-{y^2},-{y^2},0,$\\
  &$ -y,0,-{y^2},0,-y,$
  &$ -y,-y,0,-y,-y,$
 &$  0,0,0,0,0,$\\
 &$  0,0,0,0,0,$
 &$  -1,0,-y,-y,0,$
 &$  0,0,0,0,0,$\\
 &$  -{y^2},-{y^2},0,0,-{y^2},$
 &$  -{y^2},0,-y,-y,0,$
 &$  0,-y,-y,-y,-y,$\\
 &$  0,0,0,0,0,$
 &$  -{y^2},-{y^2},0,-y,0,$
 &$  0,0,-y,-y,-y,$\\
 &$  0,0,-y,-y,-y,$
 &$  0,0,0,0,0,$
 &$  0,-{y^2},0,-y,-y,$\\
 &$  -y,0,-y,-y,-y,$
 &$  0,0,0,0,0,$
 &$  -y,-y,-y,0,0,$\\
 &$  0,0,0,0,0,$
 &$  0,0,0,0,0,$
 &$  0,-y,-y,0,-{y^2},$\\
 &$  -y,-y,-y,-y,0,$
 &$  0,0,0,0,0,$
 &$  -y,0,0,-y,-y,$\\
 &$  0,-y,-y,-y,0,$
 &$  0,-y,-y,-y,0,$
 &$  0,0,0,0,0,$\\
 &$  0,0,0,0,0,$
 &$  -{y^2},0,-y,0,-{y^2} $
 &\\
\end{supertabular}
\end{center}

\newpage
\normalsize
\section{Discussions}
We have pointed out in the previous section that there exist instances
in which the elliptic genera for a pair of Landau-Ginzburg orbifolds
obey the relation (\ref{dualeg}) (or equivalently (\ref{dualchiy}))
and moreover the roles of the untwisted and twisted sectors are
exchanged. To consider this phenomenon a little bit further we shall
now concentrate on the case where two Landau-Ginzburg orbifolds are in
correspondence with sigma models as investigated in sects. 4.3 and
4.4.

Let us first extend the $\chi_y$-genus to describe the full
$U(1)\times U(1)$ charge spectrum for the $(c,c)$ ring.  For a
Calabi-Yau $\hat c$-fold ${\cal M}$ we define
\begin{equation}
  \chi[{\cal M}](y,\bar y)=\sum_{q_{\rm L},q_{\rm R}=0}^{\hat c}
  (-1)^{q_{\rm L}+q_{\rm R}}h_{q_{\rm L},\hat c-q_{\rm R}}y^{q_{\rm
    L}}\bar y^{q_{\rm R}}\,,
\label{extdgenus}
\end{equation}
where $h_{q_{\rm L},\hat c-q_{\rm R}}$ is the number of states with
charge $(q_{\rm L},q_{\rm R})$ and is also equal to the Hodge number,
$\mathop{\rm dim} H^{q_{\rm L},\hat c-q_{\rm R}}({\cal M})$.  Suppose that
a pair of Calabi-Yau $\hat c$-folds $({\cal M},\tilde{\cal M})$
consist of a mirror pair, then we have
\begin{equation}
  \tilde h_{p,q}=h_{p,\hat c-q}\,,
\end{equation}
where $\tilde h_{p,q}$ are the Hodge numbers for $\tilde{\cal M}$.
Hence the extended genus (\ref{extdgenus}) for ${\cal M}$ and
$\tilde{\cal M}$ are related through
\begin{equation}
  \chi[{\cal M}](y,\bar y)=(-1)^{\hat c}\bar y^{\hat c}\chi[\tilde{\cal
    M}](y,1/\bar y)\,.
\end{equation}
Setting $\bar y=1$ yields the relation for $\chi_y$-genera
\begin{equation}
  \chi_y[{\cal M}]=(-1)^{\hat c}\chi_y[\tilde{\cal M}]\,.
\end{equation}

When the $\hat c$-fold ${\cal M}$ has a correspondence with a
Landau-Ginzburg orbifold $W/\!/G$ one can write down $\chi[{\cal
  M}](y,\bar y)=\pm\chi[W/\!/G](y,\bar y)$ explicitly.
Following the reasoning in \cite{rVafaii}\cite{rIV}\ we find from
(\ref{chiygenus}) that
\begin{eqnarray}
 &&\chi[W/\!/G](y,\bar y)=\frac{(-1)^N}{\vert
    G\vert}\sum_{{\boldsymbol \alpha},{\boldsymbol \beta}\in G}
  \prod_{\stackrel{\scriptstyle i}{\omega_i\alpha_i\not\in{\bf Z}}}
  (y\bar y)^{\frac{1}{2}(1-2\omega_i)}(y/\bar
  y)^{-(\!(\omega_i\alpha_i)\!)}\nonumber\\
  &&\hspace{1.5cm}\times \prod_{\stackrel{\scriptstyle
      i}{\omega_i\alpha_i\in{\bf Z}}} \bfe{\omega_i\beta_i+\frac{1}{2}}
\frac{1-\bfe{(1-\omega_i)\beta_i}(y\bar y)^{1-\omega_i}}
{1-\bfe{\omega_i\beta_i}(y\bar y)^{\omega_i}}\,.
  \label{extdgenuslg}
\end{eqnarray}

We now consider a $3$-fold ${\cal M}=\hat {\cal M}_G$ where ${\cal
  M}_G$ is given by (\ref{fermat}) with $d=5$. Inspecting the tables
for $G_k\ (k=0,1,\ldots,7)$ we see that
$\sum_{\omega_i\alpha_i\not\in{\bf Z}}(\!(\omega_i\alpha_i)\!)=0$
occurs only in the untwisted sectors\footnote{This is in no way a
  generic situation.}. Thus it is clear from (\ref{extdgenuslg}) that
the states with $(q_{\rm L},q_{\rm R})=(1,1)$ corresponding to
$H^{1,2}({\cal M})$, come from the untwisted sectors while the states
with $(q_{\rm L},q_{\rm R})=(1,2)$ corresponding to $H^{1,1}({\cal
  M})$, from the twisted sectors.  Similarly, for its mirror partner
$\tilde {\cal M}$ described as $W/\!/G^*$, the elements of
$H^{1,2}(\tilde {\cal M})$ (or $ H^{1,1}(\tilde {\cal M})$) arise from
the untwisted (or twisted) sectors.  To see more explicitly let us
evaluate (\ref{extdgenuslg}) for $G=G_2$ and $G^*=G_5$.  We obtain
\begin{equation}
  \chi[W/\!/G](y,\bar y)= \chi_u[W/\!/G](y,\bar y)+
  \chi_t[W/\!/G](y,\bar y)\,,
\end{equation}
where $\chi_u$ (or $\chi_t$) stands for the contribution from the
untwisted (or twisted) sectors:
\begin{eqnarray}
 \chi_u[W/\!/G](y,\bar y)&=&1+21y\bar y+21y^2\bar y^2+y^3\bar y^3\,,\\
 \chi_t[W/\!/G](y,\bar y)&=&-y^3-y^2\bar y-y\bar y^2-\bar y^3\,.
\end{eqnarray}
For $W/\!/G^*$ we get
\begin{eqnarray}
 \chi_u[W/\!/G^*](y,\bar y)&=&1+y\bar y+y^2\bar y^2+y^3\bar y^3\,,\\
 \chi_t[W/\!/G^*](y,\bar y)&=&-y^3-21y^2\bar y-21y\bar y^2-\bar y^3\,.
\end{eqnarray}
Hence
\begin{eqnarray}
  \chi_u[W/\!/G](y,\bar y)&=&-\bar y^3\chi_t[W/\!/G^*](y,1/\bar y)\,,\\
  \chi_t[W/\!/G](y,\bar
  y)&=&-\bar y^3\chi_u[W/\!/G^*](y,1/\bar y)\,.
\end{eqnarray}
Upon setting $\bar y=1$ this reduces to the relation for
$\chi_y$-genera we found in sect 4.4. We have checked using
(\ref{extdgenuslg}) that similar results hold for all the examples
given in sect.4.4.  Therefore what we have observed seems to be
natural whenever a mirror pair has a corresponding pair of
Landau-Ginzburg orbifolds.

\footnotesize

\landscape

\begin{table}[tbp]
  \begin{center}
    \leavevmode
\begin{tabular}{lll}\hline
  $(h,d_1,d_2,d_3)$&$W(z_1,z_2,z_3)$&$\chi_y[W] \qquad(t=y^{1/h})$\\
  \hline &&\\ $(12,4,4,3)$&$z_1^3+z_2^3+z_3^4$&$1 + {t^3} + 2\,{t^4} +
  {t^6} + 2\,{t^7} + {t^8} + 2\,{t^{10}} + {t^{11}} + {t^{14}}$\\
  $(13,4,3,5)$&$z_1^2z_3+z_2z_3^2+z_1z_2^3$&$1 + {t^3} + {t^4} + {t^5}
  + {t^6} + {t^87} + {t^8} + {t^9} + {t^{10}} + {t^{11}} + {t^{12}} +
  {t^{15}}$\\ $(15,6,5,3)$&$z_1^2z_3+z_2^3+z_2^5$&$1 + {t^3} + {t^5} +
  2\,{t^6} + {t^8} + {t^9} + 2\,{t^{11}} + {t^{12}} + {t^{14}} +
  {t^{17}}$\\ $(16,5,4,6)$&$z_1^2z_3+z_2z_3^2+z_2^4$&$1 + {t^4} +
  {t^5} + {t^6} + {t^8} + {t^9} + {t^{10}} + {t^{12}} + {t^{13}} +
  {t^{14}} + {t^{18}}$\\ $(16,4,3,8)$&$z_1^4+z_1z_2^4+z_3^2$&$ 1 +
  {t^3} + {t^4} + {t^6} + {t^7} + {t^8} + {t^9} + {t^{10}} + {t^{11}}
  + {t^{12}} + {t^{14}} + {t^{15}} + {t^{18}}$\\
  $(18,7,6,4)$&$z_1^2z_3+z_2^3+z_2z_3^3$&$ 1 + {t^4} + {t^6} + {t^7} +
  {t^8} + {t^{10}} + {t^{12}} + {t^{13}} + {t^{14}} + {t^{16}} +
  {t^{20}}$\\ $(18,5,3,9)$&$z_1^3z_2+z_2^6+z_3^2$&$ 1 + {t^3} + {t^5}
  + {t^6} + {t^8} + {t^9} + {t^{10}} + {t^{11}} + {t^{12}} + {t^{14}}
  + {t^{15}} + {t^{17}} + {t^{20}}$\\
  $(20,5,4,10)$&$z_1^4+z_2^5+z_3^2$&$ 1 + {t^4} + {t^5} + {t^8} +
  {t^9} + {t^{10}} + {t^{12}} + {t^{13}} + {t^{14}} + {t^{17}} +
  {t^{18}} + {t^{22}}$\\ $(22,6,4,11)$&$z_1^3z_2+z_1z_2^4+z_3^2$&$ 1 +
  {t^4} + {t^6} + {t^8} + {t^{10}} + 2\,{t^{12}} + {t^{14}} + {t^{16}}
  + {t^{18}} + {t^{20}} + {t^{24}}$\\
  $(24,9,8,6)$&$z_1^2z_3+z_2^3+z_3^4$&$ 1 + {t^6} + {t^8} + {t^9} +
  {t^{12}} + {t^{14}} + {t^{17}} + {t^{18}} + {t^{20}} + {t^{26}}$\\
  $(24,8,3,12)$&$z_1^3+z_2^8+z_3^2$&$1 + {t^3} + {t^6} + {t^8} + {t^9}
  + {t^{11}} + {t^{12}} + {t^{14}} + {t^{15}} + {t^{17}} + {t^{18}} +
  {t^{20}} + {t^{23}} + {t^{26}}$\\
  $(30,8,6,15)$&$z_1^3z_2+z_2^5+z_3^2$&$1 + {t^6} + {t^8} + {t^{12}} +
  {t^{14}} + {t^{16}} + {t^{18}} + {t^{20}} + {t^{24}} + {t^{26}} +
  {t^{32}}$\\ $(30,10,4,15)$&$z_1^3+z_1z_2^5+z_3^2$&$ 1 + {t^4} +
  {t^8} + {t^{10}} + {t^{12}} + {t^{14}} + {t^{16}} + {t^{18}} +
  {t^{20}} + {t^{22}} + {t^{24}} + {t^{28}} + {t^{32}}$\\
  $(42,14,6,21)$&$z_1^3+z_2^7+z_3^2$&$ 1 + {t^6} + {t^{12}} + {t^{14}}
  + {t^{18}} + {t^{20}} + {t^{24}} + {t^{26}} + {t^{30}} + {t^{32}} +
  {t^{38}} + {t^{44}}$ \\ &&\\ \hline
    \end{tabular}
  \end{center}
  \caption{Exceptional singularities:  $\chi_y$-genera.}
  \label{tab:tabA}
\end{table}

\begin{table}[tbp]
  \begin{center}
    \leavevmode
\begin{tabular}{rl}\hline
$(h,d_1,d_2,d_3)$&$\{-\chi_y^0[W/\!/G_0],-\chi_y^1[W/\!/G_0],\ldots,
-\chi_y^{h-1}[W/\!/G_0]\}$\\
 \hline
&\\
$(12,4,4,3)$ & $\{ 0,{t^{14}},{t^3},2\,{t^4},0,{t^6},2\,{t^7},{t^8},0,2\,%
{t^{10}},{t^{11}},1\}$\\
$(13,4,3,5)$ & $\{ 0,{t^{15}},{t^3},{t^4},{t^5},{t^6},{t^7},{t^8},{t^9},%
{t^{10}},{t^{11}},{t^{12}},1\}$\\
$(15,6,5,3)$ & $\{ 0,{t^{17}},{t^3},0,{t^5},2\,{t^6},0,{t^8},{t^9},0,2\,%
{t^{11}},{t^{12}},0,{t^{14}},1\}$\\
$(16,5,4,6)$ & $\{ 0,{t^{18}},{t^3},{t^4},0,{t^6},{t^7},{t^8},{t^9},%
{t^{10}},{t^{11}},{t^{12}},0,{t^{14}},{t^{15}},1\}$\\
$(16,4,3,8)$ & $\{ 0,{t^{18}},0,{t^4},{t^5},{t^6},0,{t^8},{t^9},%
{t^{10}},0,{t^{12}},{t^{13}},{t^{14}},0,1\}$\\
$(18,7,6,4)$ & $\{ 0,{t^{20}},{t^3},0,{t^5},{t^6},0,{t^8},{t^9},{t^{10}},%
{t^{11}},{t^{12}},0,{t^{14}},{t^{15}},0,{t^{17}},1\}$\\
$(18,5,3,9)$ & $\{ 0,{t^{20}},0,{t^4},0,{t^6},{t^7},{t^8},0,{t^{10}},0,%
{t^{12}},{t^{13}},{t^{14}},0,{t^{16}},0,1\}$\\
$(20,5,4,10)$ & $\{ 0,{t^{22}},0,{t^4},{t^5},0,0,{t^8},{t^9},{t^{10}},0,%
{t^{12}},{t^{13}},{t^{14}},0,0,{t^{17}},{t^{18}},0,1\}$\\
$(22,6,4,11)$ & $\{ 0,{t^{24}},0,{t^4},0,{t^6},0,{t^8},0,{t^{10}},0,2\,%
{t^{12}},0,{t^{14}},0,{t^{16}},0,{t^{18}},0,{t^{20}},0,1\}$\\
$(24,9,8,6)$ & $\{ 0,{t^{26}},{t^3},0,0,{t^6},0,{t^8},{t^9},0,{t^{11}},%
{t^{12}},0,{t^{14}},{t^{15}},0,{t^{17}},{t^{18}},0,{t^{20}},0,0,%
{t^{23}},1\}$\\
$(24,8,3,12)$ & $\{ 0,{t^{26}},0,0,0,{t^6},0,{t^8},{t^9},0,0,{t^{12}},0,%
{t^{14}},0,0,{t^{17}},{t^{18}},0,{t^{20}},0,0,0,1\}$\\
$(30,8,6,15)$ & $\{ 0,{t^{32}},0,{t^4},0,0,0,{t^8},0,{t^{10}},0,{t^{12}},%
0,{t^{14}},0,{t^{16}},0,{t^{18}},0,{t^{20}},0,{t^{22}},0,{t^{24}},0,0,0,%
{t^{28}},0,1\}$\\
$(30,10,4,15)$ & $\{ 0,{t^{32}},0,0,0,{t^6},0,{t^8},0,0,0,{t^{12}},0,%
{t^{14}},0,{t^{16}},0,{t^{18}},0,{t^{20}},0,0,0,{t^{24}},0,%
{t^{26}},0,0,0,1\}$\\
$(42,14,6,21)$ & $\{ 0,{t^{44}},0,0,0,{t^6},0,0,0,0,0,{t^{12}},0,%
{t^{14}},0,0,0,{t^{18}},0,{t^{20}},0,0,0,{t^{24}},0,{t^{26}},0,0,0,%
{t^{30}},0,{t^{32}},0,0,0,0,0,{t^{38}},0,0,0,1\}$\\
&\\
\hline
\end{tabular}
  \end{center}
  \caption{Exceptional singularities:the untwisted and
    twisted sector contributions to
    the Landau-Ginzburg orbifold $\chi_y$-genera.}
  \label{tab:tabB}
\end{table}

\begin{table}[tbp]
  \begin{center}
    \leavevmode
\begin{tabular}{rl}\hline
$(h,d_1,d_2,d_3)$ &
$\{\chi_y^0[W/\!/G_0],\chi_y^1[W/\!/G_0],\ldots,\chi_y^{h-1}
[W/\!/G_0]\}$\\
 \hline
&\\
$(12,4,4,3)$&$
\{ 1 + 10y + {y^2},{y^2},y,2y,0,y,2y,y,0,2y,y,1\}$\\
$(13,4,3,5)$&$
\{ 1 + 10y + {y^2},{y^2},y,y,y,y,y,y,y,y,y,y,1\}$\\
$(15,6,5,3)$&$
\{ 1 + 10y + {y^2},{y^2},y,0,y,2y,0,y,y,0,2y,y,0,y,1\}$\\
$(16,5,4,6)$&$
\{ 1 + 9y + {y^2},{y^2},y,y,0,y,y,y,y,y,y,y,0,y,y,1\}$\\
$(16,4,3,8)$&$
\{ 1 + 11y + {y^2},{y^2},0,y,y,y,0,y,y,y,0,y,y,y,0,1\}$\\
$(18,7,6,4)$&$
\{ 1 + 9y + {y^2},{y^2},y,0,y,y,0,y,y,y,y,y,0,y,y,0,y,1\}$\\
$(18,5,3,9)$&$
\{ 1 + 11y + {y^2},{y^2},0,y,0,y,y,y,0,y,0,y,y,y,0,y,0,1\}$\\
$(20,5,4,10)$&$
\{ 1 + 10y + {y^2},{y^2},0,y,y,0,0,y,y,y,0,y,y,y,0,0,y,y,0,1\}$\\
$(22,6,4,11)$&$
\{ 1 + 10y + {y^2},{y^2},0,y,0,y,0,y,0,y,0,2y,0,y,0,y,0,y,0,y,0,1\}$\\
$(24,9,8,6)$&$
\{ 1 + 8y + {y^2},{y^2},y,0,0,y,0,y,y,0,y,y,0,y,y,0,y,y,0,y,0,0,y,1\}$\\
$(24,8,3,12)$&$
\{ 1 + 12y + {y^2},{y^2},0,0,0,y,0,y,y,0,0,y,0,y,0,0,y,y,0,y,0,0,0,1\}$\\
$(30,8,6,15)$&$
\{ 1 + 9y + {y^2},{y^2},0,y,0,0,0,y,0,y,0,y,0,y,0,y,0,y,0,y,0,
y,0,y,0,0,0,y,0,1\}$\\
$(30,10,4,15)$&$
\{ 1 + 11y + {y^2},{y^2},0,0,0,y,0,y,0,0,0,y,0,y,0,y,0,y,0,y,0,0,0,y,0,
y,0,0,0,1\}$\\
$(42,14,6,21)$&$
\{ 1 + 10y + {y^2},{y^2},0,0,0,y,0,0,0,0,0,y,0,y,0,0,0,y,0,y,0,0,0,y,0,
y,0,0,0,y,0,y,0,0,0,0,0,y,0,0,0,1\}$ \\
&\\
\hline
\end{tabular}
  \end{center}
  \caption{The $K3$ associated with exceptional singularities:
the  untwisted and twisted sector contributions to the Landau-Ginzburg
 orbifold $\chi_y$-genera.}
  \label{tab:tableC}
\end{table}

\end{document}